\documentclass[reprint,twocolumn]{revtex4-2}
\usepackage{amsfonts}
\usepackage{amsmath}
\usepackage{amssymb}
\usepackage{charter}
\usepackage{graphicx}
\usepackage{float}
\usepackage{amsmath}
\usepackage{amssymb}
\usepackage{charter}
\usepackage{graphicx}
\usepackage{subfigure}
\usepackage{graphicx}
\usepackage{epstopdf}
\usepackage{color}
\usepackage{tocvsec2}
\usepackage{enumerate}
\usepackage{graphicx}
\usepackage{dcolumn}
\usepackage{bm}

\begin{document}

\title{Phase estimation via multi-photon subtraction inside the SU(1,1)
interferometer }
\author{Qingqian Kang$^{1,2}$}
\thanks{These authors contributed equally to this work and should be
considered co-fist authors.}
\author{Zekun Zhao$^{1}$}
\thanks{These authors contributed equally to this work and should be
considered co-fist authors.}
\author{Youke Xu$^{1}$}
\author{Teng Zhao$^{1}$}
\author{Cunjin Liu$^{1}$}
\author{Liyun Hu$^{1,3}$}
\thanks{Corresponding author. hlyun@jxnu.edu.cn}
\affiliation{$^{{\small 1}}$\textit{Center for Quantum Science and Technology, Jiangxi
Normal University, Nanchang 330022, China}\\
$^{{\small 2}}$\textit{Department of Physics, Jiangxi Normal University Science and
Technology College, Nanchang 330022, China}\\
$^{{\small 3}}$\textit{Institute for Military-Civilian Integration of Jiangxi Province,
 Nanchang 330200, China}}

\begin{abstract}
To improve the phase sensitivity, multi-photon subtraction schemes within
the SU(1,1) interferometer are proposed. The input states are the coherent
state and the vacuum state, and the detection method is homodyne detection.
The effects of multi-photon subtraction on phase sensitivity, quantum Fisher
information, and quantum Cram\'{e}r-Rao bound are analyzed under both ideal
and photon losses situations. It is shown that the internal subtraction
operation can improve the phase sensitivity, which becomes better
performance by increasing subtraction number. It can also efficiently
improve the robustness of the SU(1,1) interferometer against internal photon
losses. By comparing separatively arbitrary photon subtraction on the
two-mode inside SU(1,1) interferometer, the performance differences under
different conditions are analyzed, including the asymmetric properties of
non-Gaussian operations on the phase precision and the quantum Fisher
information. Our proposed scheme represents a valuable method for
achieving quantum precision measurements.

\textbf{PACS: }03.67.-a, 05.30.-d, 42.50,Dv, 03.65.Wj
\end{abstract}

\maketitle

\section{Introduction}

Quantum precision measurement, leveraging the unique properties of quantum
states, offers a highly accurate approach that has the potential for
widespread utilization in diverse fields \cite{01,02,03,04,05,06,07,08,09},
such as quantum computing and quantum communication. Remarkable achievements
in quantum measurement technologies have been made, including simultaneous
measurement and coherent measurement using quantum entanglement states.
These techniques have found applications in biology, physics, and chemistry,
enabling improved accuracy in areas such as microbial detection and drug
discovery, surpassing classical measurement techniques. Moreover, studying
quantum measurement technology not only deepens our understanding of quantum
mechanics but also profoundly influences the development of quantum
information science. Recognizing that the measurement process itself can
introduce disturbances to quantum states, it is essential to identify and
mitigate sources of interference, and develop more robust interference
cancellation techniques. As the demand for computation and communication
continues to grow, the application of quantum measurement technology will
expand across various domains. Consequently, the research on quantum
precision measurement holds immense value, as it opens up endless
opportunities for scientific exploration and technological advancement.

Phase estimation is a critical aspect of precision measurement, with optical
interferometers playing a key role in this technique. In 1986, Yurke \emph{%
et al}. \cite{a1} introduced the SU(1,1) interferometer, which replaced
traditional beam splitters (BSs) with optical parametric amplifiers (OPAs).
The entangled state generated by the first OPA leads to noiseless
amplification during the quantum destructive interference of the SU(1,1)
interferometer \cite{m1}, enhancing the precision of phase estimation. This
technique revolutionized phase estimation, becoming a vital tool in quantum
precision measurements. By utilizing entangled photon states, the SU(1,1)
interferometer can surpass the standard quantum limit (SQL), enabling higher
precision. In recent years, there has been significant interest in studying
the SU(1,1) interferometer \cite{m3}. In 2011, Jing \emph{et al}. \cite{2}
successfully implemented this interferometer experimentally. In this nonlinear interferometer,
the maximum output intensity can be much
higher than that of linear interferometer due to the OPA.  Apart from the
standard form, various configurations of SU(1,1) interferometer have also
been proposed \cite{m4,m5,m6,m7,m9,m10,m11,m12}.

Photon loss inevitably decreases the phase sensitivity during the estimation
process, presenting a significant challenge in mitigating its adverse
effects. In 2012, Marino \emph{et al}. \cite{3} investigated the impact of
losses on the phase sensitivity of the SU(1,1) interferometer with intensity
detection. Their findings revealed that although propagation losses reduced
phase sensitivity, it was still feasible to surpass the SQL even in the
presence of substantial losses. Subsequently, in 2014, Li \emph{et al}. \cite%
{4} demonstrated that the SU(1,1) interferometer employing coherent and
squeezed-vacuum states can approach the Heisenberg Limit (HL) through
homodyne detection. Additionally, Manceau \emph{et al}. \cite{5} illustrated
that increasing the gain of the second amplifier enables the interferometer
to maintain phase supersensitivity despite up to 80\% detection losses.

To further fulfill the aforementioned goal, effective methods grounded in
non-Gaussian operations, such as photon subtraction \cite{m13,m14,m15}, have
been proposed. These operations are notably important in quantum
communication and quantum computation \cite{10,a21,a22,a23}. They present
the potential to generate a more strongly entangled sub-ensemble from a
weakly entangled state \cite{m16} and contribute to enhancing quantum
measurement precision \cite{m17,m18,m19}. Experimental studies have
illustrated the conditional generation of superpositions of distinct quantum
operations through single-photon interference, providing a practical
approach for preparing non-Gaussian operations \cite{m20}. This advancement
has unveiled new possibilities in quantum state manipulation and
implications for various quantum technologies.

As previously mentioned, although SU(1,1) interferometer is
highly valuable for precision measurement \cite{b1,b2}, the precision is
still affected by dissipation, particularly photon losses inside the
interferometer \cite{3,7}. Consequently, to further enhance precision, non-Gaussian operations should
serve as an alternative approach to mitigate internal dissipation. In
Reference \cite{9}, single-photon subtraction within the SU(1,1)
interferometer is utilized to enhance phase sensitivity, thereby increasing
robustness against internal losses. Notably, this process only involves the
simultaneous subtraction of single photons from two modes.

In this paper, our focus lies in enhancing precision by individually
performing arbitrary photon subtraction on the two-mode inside the SU(1,1)
interferometer to comprehend the effects of non-Gaussian operations. This
includes analyzing the asymmetric properties of non-Gaussian operations on
phase precision and the quantum Fisher information (QFI). For instance, it is
to understand on which mode multi-photon subtraction
scheme works better. The paper is structured as follows. Sec.
II outlines the theoretical model of the multi-PSS. Sec. III delves into
phase sensitivity, encompassing both the ideal case and the internal photon
losses case. Sec. IV centers on the QFI and quantum Cram\'{e}r-Rao bound
(QCRB) \cite{b3,b4}. Finally, Sec. V provides a comprehensive summary.

\section{Model}

In this section, we commence by introducing the SU(1,1) interferometer,
depicted in Fig. 1(a). Typically, the SU(1,1) interferometer comprises two
optical parametric amplifiers (OPAs) and a linear phase shifter,
representing one of the most commonly employed interferometers in quantum
metrology research. The first OPA is characterized by a two-mode squeezing
operator $U_{S_{1}}(\xi )=\exp (\xi _{1}^{\ast }ab-\xi _{1}a^{\dagger
}b^{\dagger })$, where $a$ ($b$), $a^{\dagger }$ ($b^{\dagger }$) represent the
photon annihilation and photon creation operators, respectively. The squeezing
parameter $\xi _{1}=g_{1}e^{i\theta _{1}}$ can be expressed in terms of a
gain factor $g_{1}$ and a phase shift $\theta _{1}$, and plays a critical
role in shaping the interference pattern and determining the phase
sensitivity of the system. Following the first OPA, mode $a$ undergoes a
phase shift process $U_{\phi }=\exp [i\phi (a^{\dagger }a)]$, while mode $b$
remains unchanged. Subsequently, the two beams are coupled in the second OPA
with the operator $U_{S_{2}}(\xi )=\exp (\xi _{2}^{\ast }ab-\xi
_{2}a^{\dagger }b^{\dagger })$, where $\xi _{2}=g_{2}e^{i\theta _{2}}$ and $%
\theta _{2}-\theta _{1}=\pi $. In the proposed scheme, $g_{1}=g_{2}=g$, and
we utilize coherent state $\left \vert \alpha \right \rangle _{a}$ and
vacuum state $\left \vert 0\right \rangle _{b}$ as input states. Homodyne
detection on the $a$-mode of the output is employed.

The SU(1,1) interferometer is generally susceptible to photon losses,
particularly in the case of internal losses. To simulate photon losses, the
use of fictitious BS is proposed, as depicted in Fig. 1(a). The operators of
these fictitious BSs can be described as $U_{B}=U_{B_{a}}\otimes U_{B_{b}}$,
with $U_{B_{a}}=\exp \left[ \theta _{a}\left( a^{\dagger
}a_{v}-aa_{v}^{\dagger }\right) \right] $ and $U_{B_{b}}=\exp \left[ \theta
_{b}\left( b^{\dagger }b_{v}-bb_{v}^{\dagger }\right) \right] $, where $%
a_{v} $ and $b_{v}$ represent vacuum modes. Here, $T_{k}$ (where ($k=a,b$))
denotes the transmissivity of the fictitious BSs, associated with $\theta
_{k}$ through $T_{k}=\cos ^{2}\theta _{k}\in \left[ 0,1\right] $. A
transmissivity value of ($T_{k}=1$) corresponds to the ideal case without
photon losses \cite{c0}. In an expanded space, the expression for the output
state of the standard SU(1,1) interferometer can be represented as the
following pure state, i.e.,
\begin{equation}
\left \vert \Psi _{out}^{0}\right \rangle =U_{S_{2}}U_{\phi
}U_{B}U_{S_{1}}\left \vert \alpha \right \rangle _{a}\left \vert 0\right
\rangle _{b}\left \vert 0\right \rangle _{a_{v}}\left \vert 0\right \rangle
_{b_{v}}.  \label{a1}
\end{equation}%
\begin{figure}[tph]
\label{Figure1} \centering \includegraphics[width=0.9%
\columnwidth]{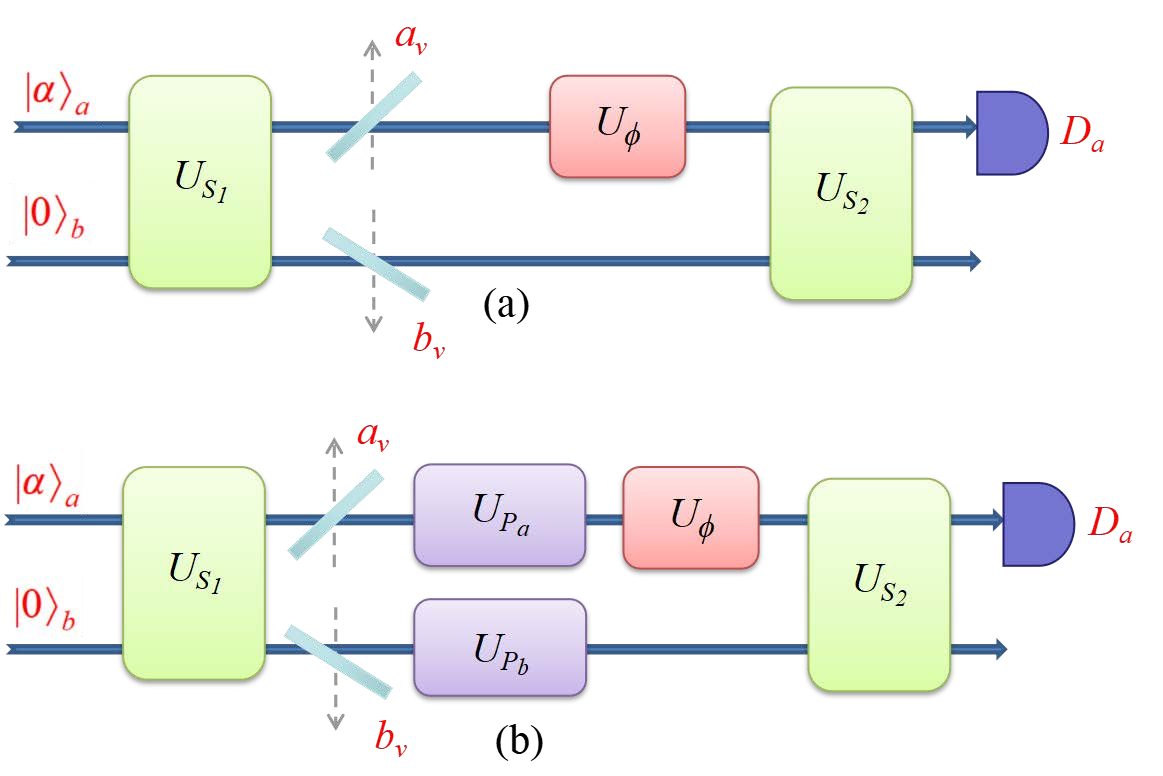} \
\caption{Schematic diagram of the SU(1,1) interferometer. (a) the standard
SU(1,1) interferometer, (b) the SU(1,1) interferometer with multi-PSS. The
two input ports are a coherent state $\left \vert \protect \alpha \right
\rangle _{a}$ and a vacuum state $\left \vert 0\right \rangle _{b}$. $a_{v}$
and $b_{v}$ are vacuum modes. $U_{S_{1}}$ and $U_{S_{2}}$\ are the optical
parametric amplifier, $U_{\protect \phi }$ is the phase shifter. $U_{P}$ is
the multi-photon subtraction operation and $D_{a}$ is the homodyne detector.}
\end{figure}

To mitigate the impact of photon losses, following the first OPA, we
introduce the multi-PSS on the two modes within the SU(1,1) interferometer,
illustrated in Fig. 1(b). In the multi-PSS, $m$ and $n$ photons are
subtracted from mode $a$ and mode $b$, respectively. This process can be
described by the operator $U_{P}=a^{m}\otimes b^{n}$. Consequently, the
output state of the interferometer can be expressed in the following form
\begin{equation}
\left \vert \Psi _{out}^{1}\right \rangle =AU_{S_{2}}U_{\phi
}U_{P}U_{B}U_{S_{1}}\left \vert \alpha \right \rangle _{a}\left \vert
0\right \rangle _{b}\left \vert 0\right \rangle _{a_{v}}\left \vert 0\right
\rangle _{b_{v}}.  \label{a2}
\end{equation}%
The normalization constant for the multi-PSS, denoted by $A$, is given by
\cite{c0}
\begin{equation}
A^{-2}=D_{m_{1},n_{1,}m_{2},n_{2}}e^{w_{1}},  \label{a3}
\end{equation}%
where $D_{m_{1},n_{1,}m_{2},n_{2}}=\partial ^{m_{1}+n_{1}+m_{2}+n_{2}}$/$%
\partial \lambda _{1}^{m_{1}}\partial \lambda _{2}^{n_{1}}\partial \lambda
_{3}^{m_{2}}\partial \lambda _{4}^{n_{2}}$ $\left. \left( \cdot \right)
\right \vert _{\lambda _{1}=\lambda _{2}=\lambda _{3}=\lambda _{4}=0},$ and%
\begin{eqnarray}
w_{1} &=&u_{1}u_{2}+u_{3}u_{4}+u_{3}\alpha \\
&&+\left( \lambda _{1}\sqrt{T_{k}}\cosh g+u_{4}\right) \alpha ^{\ast },
\label{a4} \\
u_{1} &=&-\lambda _{1}\sqrt{T_{k}}e^{-i\theta _{1}}\sinh g,  \label{a5} \\
u_{2} &=&\sqrt{T_{k}}\left( \lambda _{2}\cosh g-\lambda _{3}e^{i\theta
_{1}}\sinh g\right) ,  \label{a6} \\
u_{3} &=&\sqrt{T_{k}}\left( \lambda _{3}\cosh g-\lambda _{2}e^{-i\theta
_{1}}\sinh g\right) ,  \label{7} \\
u_{4} &=&-\lambda _{4}\sqrt{T_{k}}e^{i\theta _{1}}\sinh g.  \label{a8}
\end{eqnarray}

\section{Phase sensitivity}

Quantum metrology, an effective approach utilizing quantum resources for precise phase
measurements \cite{c1,c2}, aims to achieve highly sensitive measurements of
unknown phases. In this section, we further investigate the phase sensitivity for the
multi-PSS within the SU(1,1) interferometer \cite{c3}. Various detection methods are available for this purpose,
such as homodyne detection \cite{b5,b6}, parity detection \cite{b7,b8}, and
intensity detection \cite{b10}. Each of these methods offers different
trade-offs between sensitivity, complexity, and practical implementation. It
is important to note that for different input states and interferometers,
the phase sensitivities of various detection schemes may be different \cite%
{b11}.

Here, we take homodyne detection as the detection method of the output $a$
due to the fact that it is often the most straightforward method to
implement experimentally. In homodyne detection, the measured variable is is
one of the two orthogonal components of the mode $a$, i.e., $X=(a+a^{\dagger
})/\sqrt{2}.$ According to the error propagation equation \cite{a1}, the
phase sensitivity can be written as
\begin{equation}
\Delta \phi =\frac{\sqrt{\left \langle \Delta ^{2}X\right \rangle }}{%
|\partial \left \langle X\right \rangle /\partial \phi |}=\frac{\sqrt{\left
\langle X^{2}\right \rangle -\left \langle X\right \rangle ^{2}}}{|\partial
\left \langle X\right \rangle /\partial \phi |}.  \label{a10}
\end{equation}%
Based on Eqs. (\ref{a2}) and (\ref{a10}), the phase sensitivity for the
multi-PSS can be theoretically determined. The detail calculation steps for
the phase sensitivity $\Delta \phi $ of the multi-PSS are provided in
Appendix A.

\subsection{Ideal case}

First, we explore the ideal case, corresponding to $T_{k}=1$ (where ($k=a,b$%
)), i.e., without photon losses. For different numbers of photons
subtracted, we depict the phase sensitivity $\Delta \phi $ as a function of $%
\phi $ in Fig. 2, including single-mode photon subtraction (Fig. 2(b)),
as well as symmetrical and asymmetrical two-mode photon subtraction (Fig.
2(a) and (c)). The observations derived from Fig. 2 are as follows. (i) The
phase sensitivity initially improves and then diminishes as the phase
increases, with the optimal sensitivity deviating from $\phi =0$. (ii) It is
noteworthy that photon subtraction within the SU(1,1) interferometer
effectively enhances the phase sensitivity $\Delta \phi $, particularly with
an increased number of subtracted photons on both modes. (iii) In the case
of single-mode photon subtraction, it is evident that the performance of
photon subtraction from mode $b$ surpasses that from mode $a$ at small phase
values, while the reverse holds true at larger phase values (see Fig. 2(b)).
(iv) Similarly, this observation applies to asymmetrical two-mode photon
subtraction (refer to Fig. 2(c)), indicating the asymmetric impact of photon
subtraction on modes $a$ and $b$, with the optimal sensitivity achieved
through photon subtraction from mode $a$.
\begin{figure}[tph]
\label{Figure2} \centering
\subfigure{
\begin{minipage}[b]{0.5\textwidth}
\includegraphics[width=0.83\textwidth]{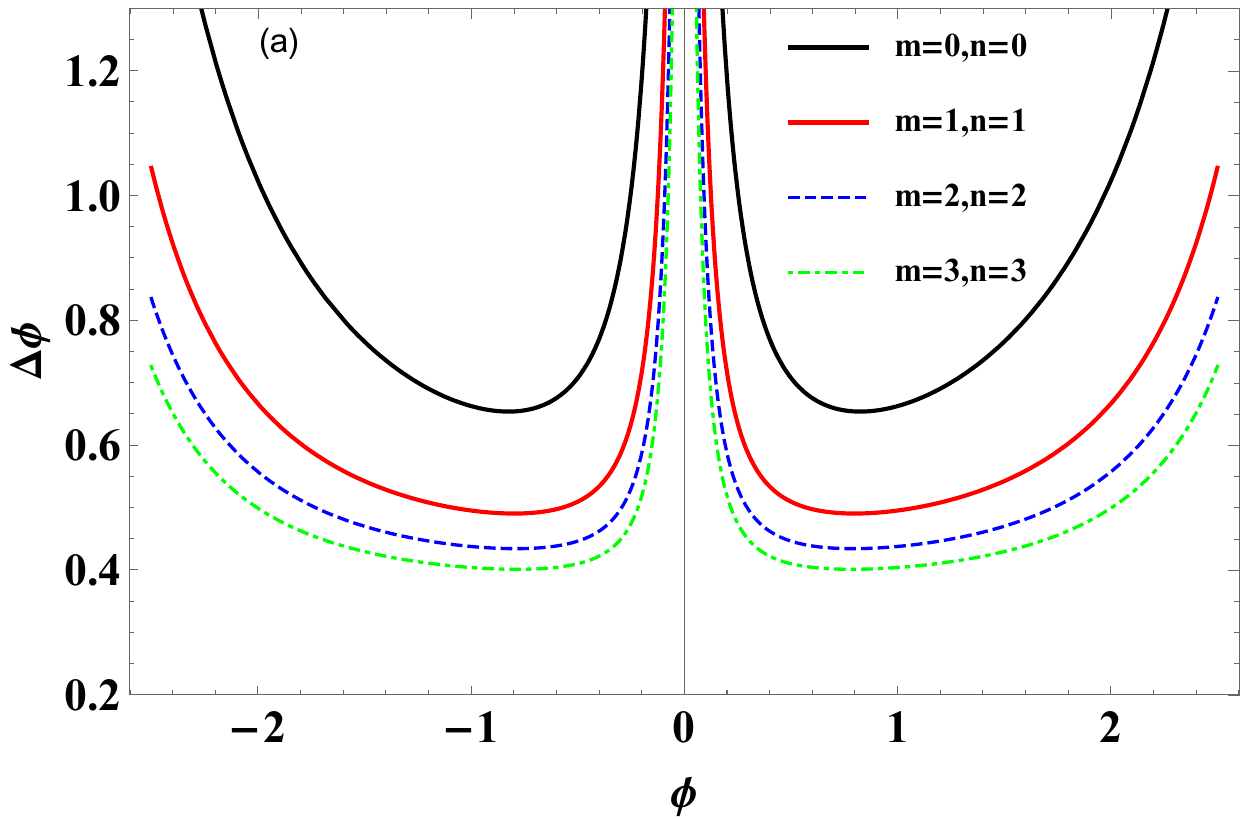}\\
\includegraphics[width=0.83\textwidth]{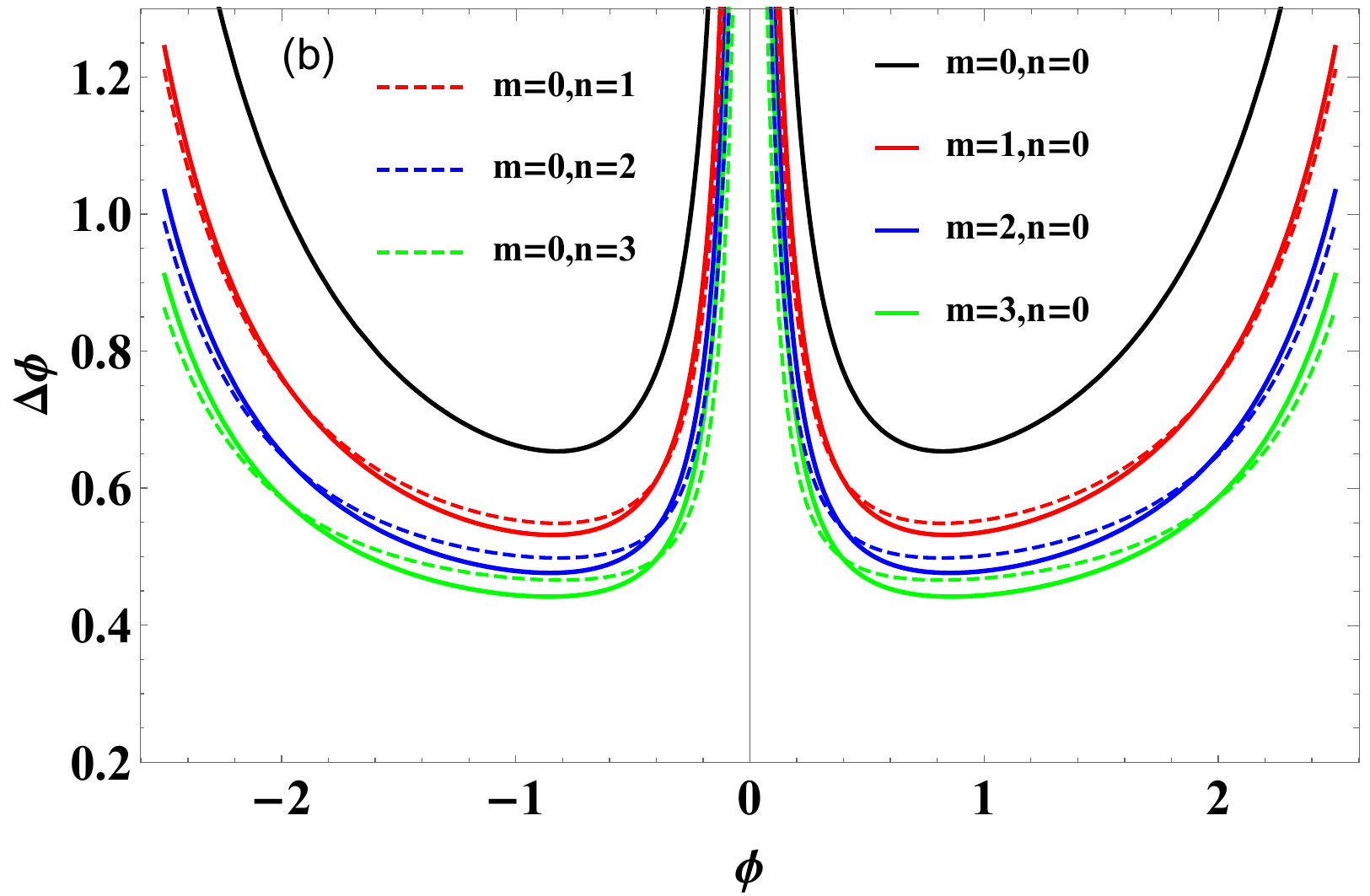}\\
\includegraphics[width=0.83\textwidth]{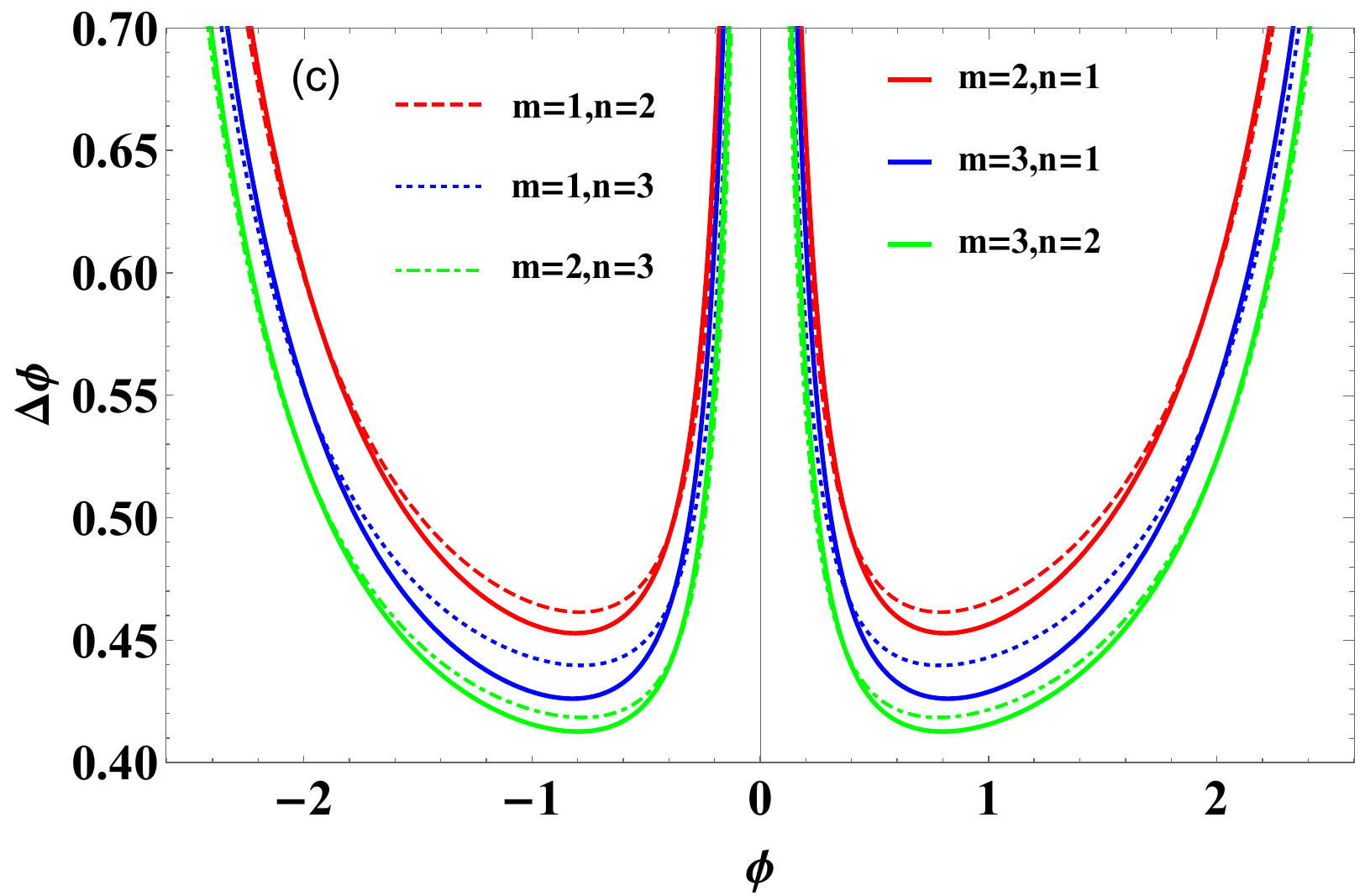}
\end{minipage}}
\caption{The phase sensitivity of multi-PSS based on the homodyne detection
as a function of $\protect \phi $ with $\protect \alpha =1$ and $g=1$. (a)
symmetrical two-mode multi-PSS, (b) single mode multi-PSS, (c) Arbitrary multi-PSS.}
\end{figure}

In Fig. 3, the phase sensitivity $\Delta \phi $ is plotted as a function of
the gain factor $g$ for different numbers of subtracted photons. The plot
demonstrates that the phase sensitivity is enhanced with an increase in the
gain factor $g$, and this improvement is further enhanced with an increase
in the number of subtracted photons. Particularly noteworthy from Fig. 3(b)
is the observation that the multi-PSS on mode $b$ exhibits higher phase
sensitivity than that on mode $a$ when the value of $g$ is small, while the
reverse is true when the value of $g$ is large. Furthermore, in the case of
asymmetrical multi-PSS on mode $b$, the change in phase sensitivity is
relatively flat with increasing $g$. Conversely, for the asymmetrical
multi-PSS on mode $a$, the enhancement in phase sensitivity initially
 improves and then diminishes with the gain factor $g$.
\begin{figure}[tph]
\label{Figure3} \centering
\subfigure{
\begin{minipage}[b]{0.5\textwidth}
\includegraphics[width=0.83\textwidth]{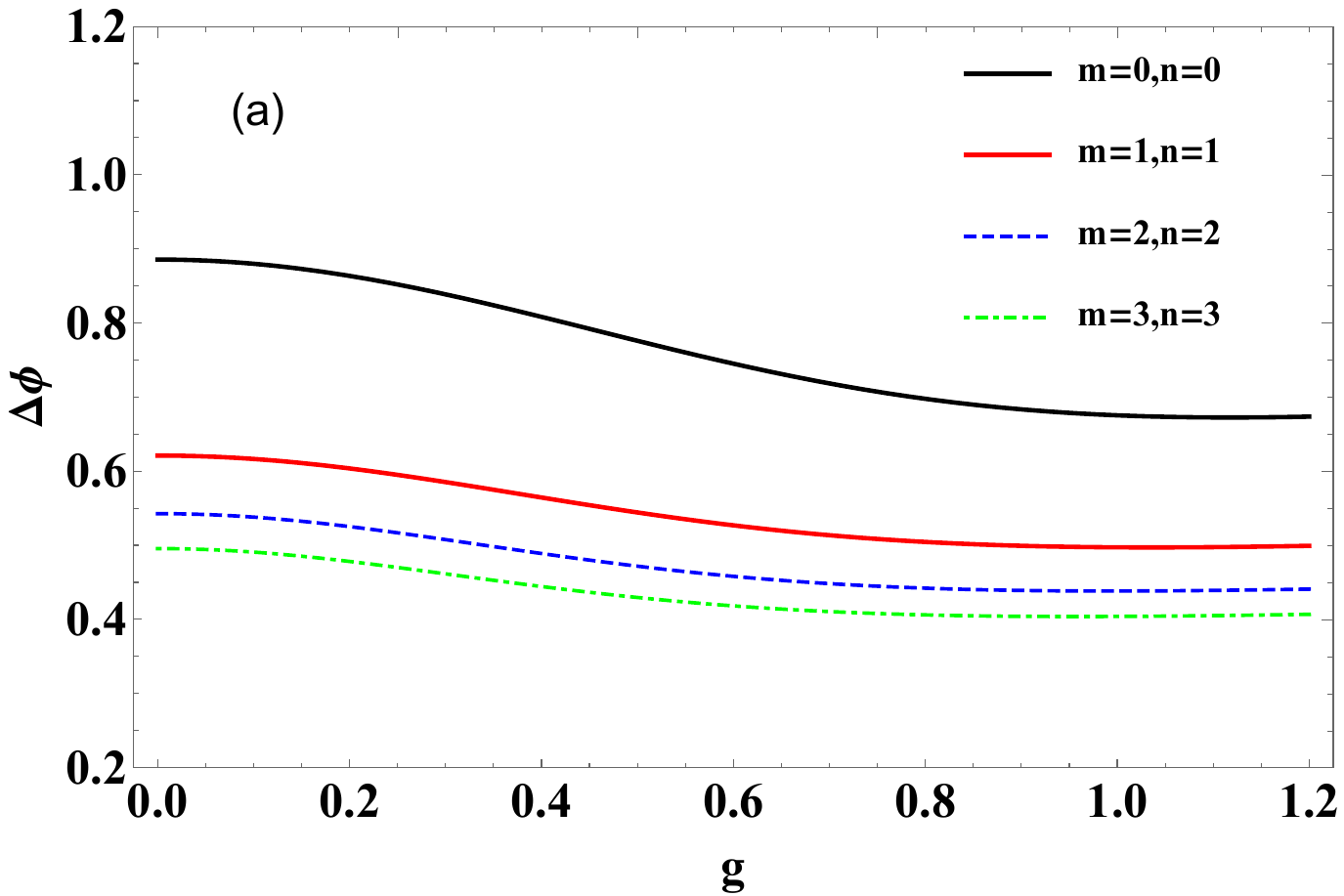}\\
\includegraphics[width=0.83\textwidth]{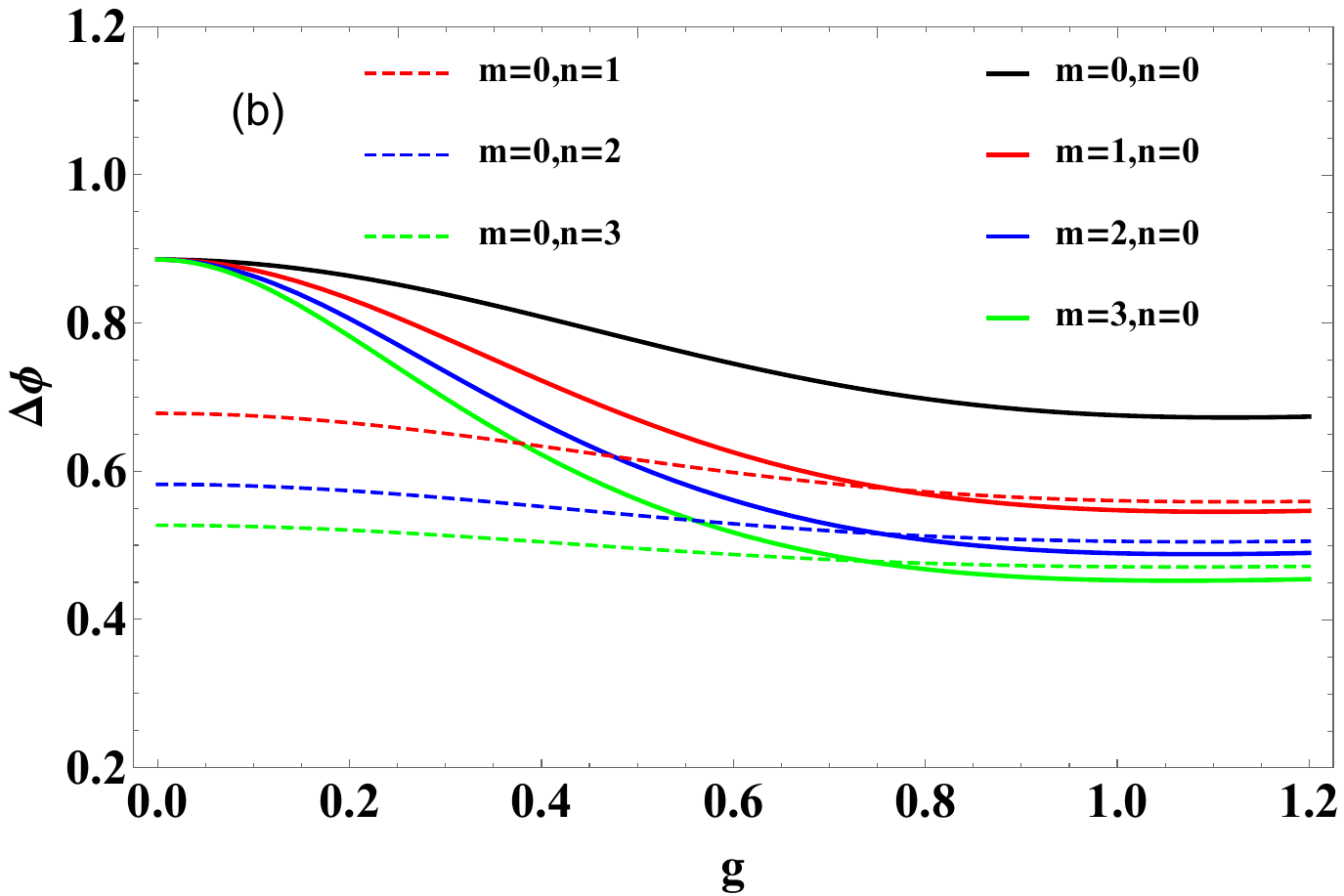}
\end{minipage}}
\caption{The phase sensitivity as a function of $g$, with $\protect \alpha =1$
and $\protect \phi =0.6$. (a) symmetrical two-mode multi-PSS, (b) single mode multi-PSS.}
\end{figure}

Similarly, we examine the phase sensitivity $\Delta \phi $ as a function of
the coherent amplitude $\alpha $, as illustrated in Fig. 4. Several similar
findings are observed compared to those in Fig. 3. For example, the phase
sensitivity exhibits improvement with the coherent amplitude $\alpha $,
which can be attributed to an increase in the average photon number with $%
\alpha $. Moreover, the enhancement effect initially grows and then
diminishes with the coherent amplitude $\alpha $ and the number of
subtracted photons. Once again, the asymmetrical property of multi-PSS
on modes $a$ and $b$ is evident. Notably, as depicted in Fig.
4(b), the multi-PSS on mode $a$ yields higher phase sensitivity than that on
mode $b$ in a smaller region of $\alpha $, whereas the converse holds true
in a larger region of $\alpha $. These findings indicate that the selection
of the mode for implementing the multi-PSS depends on the specific
parameters and requirements of the measurement task.
\begin{figure}[tbh]
\label{Figure4} \centering%
\subfigure{
\begin{minipage}[b]{0.5\textwidth}
\includegraphics[width=0.83\textwidth]{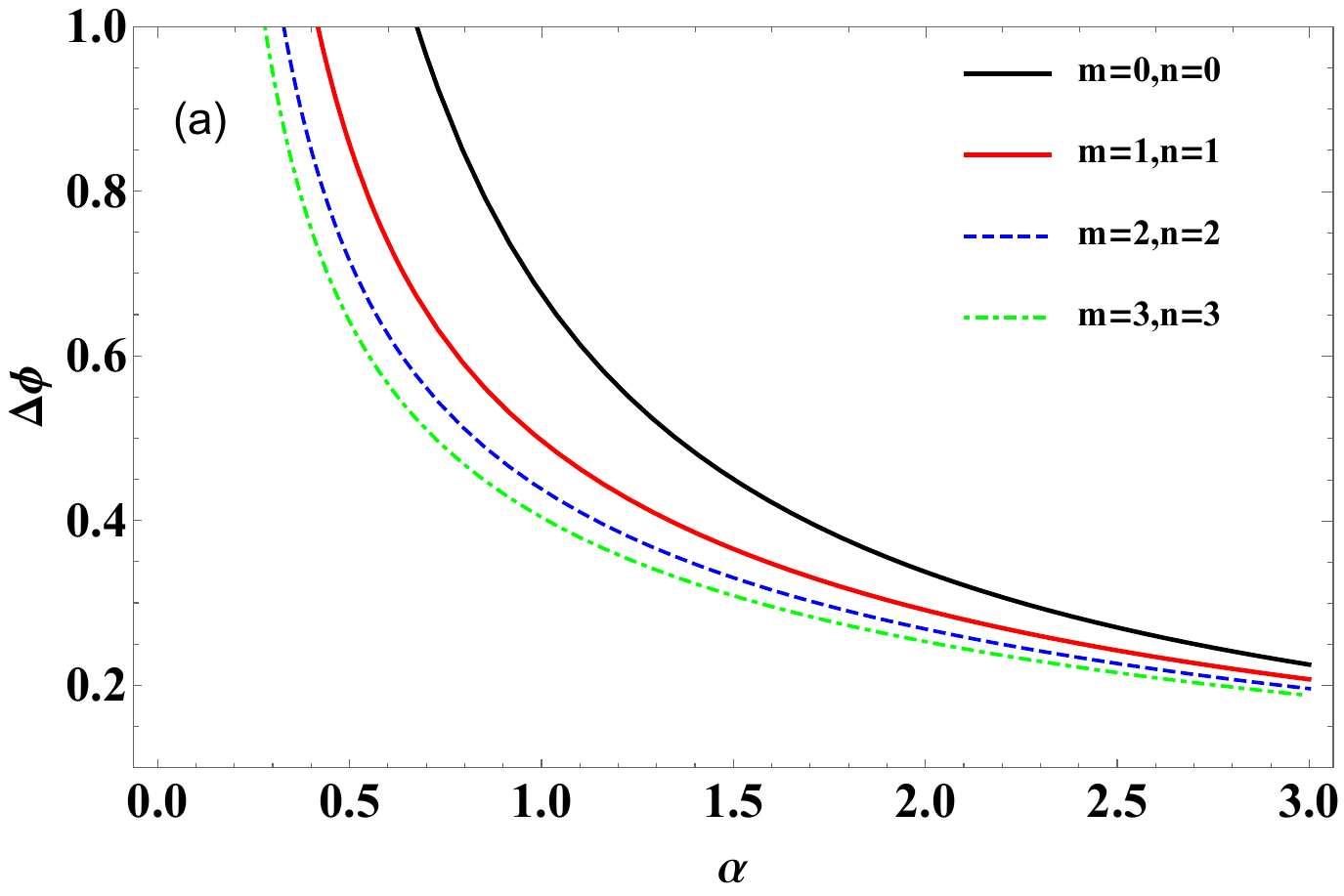}\\
\includegraphics[width=0.83\textwidth]{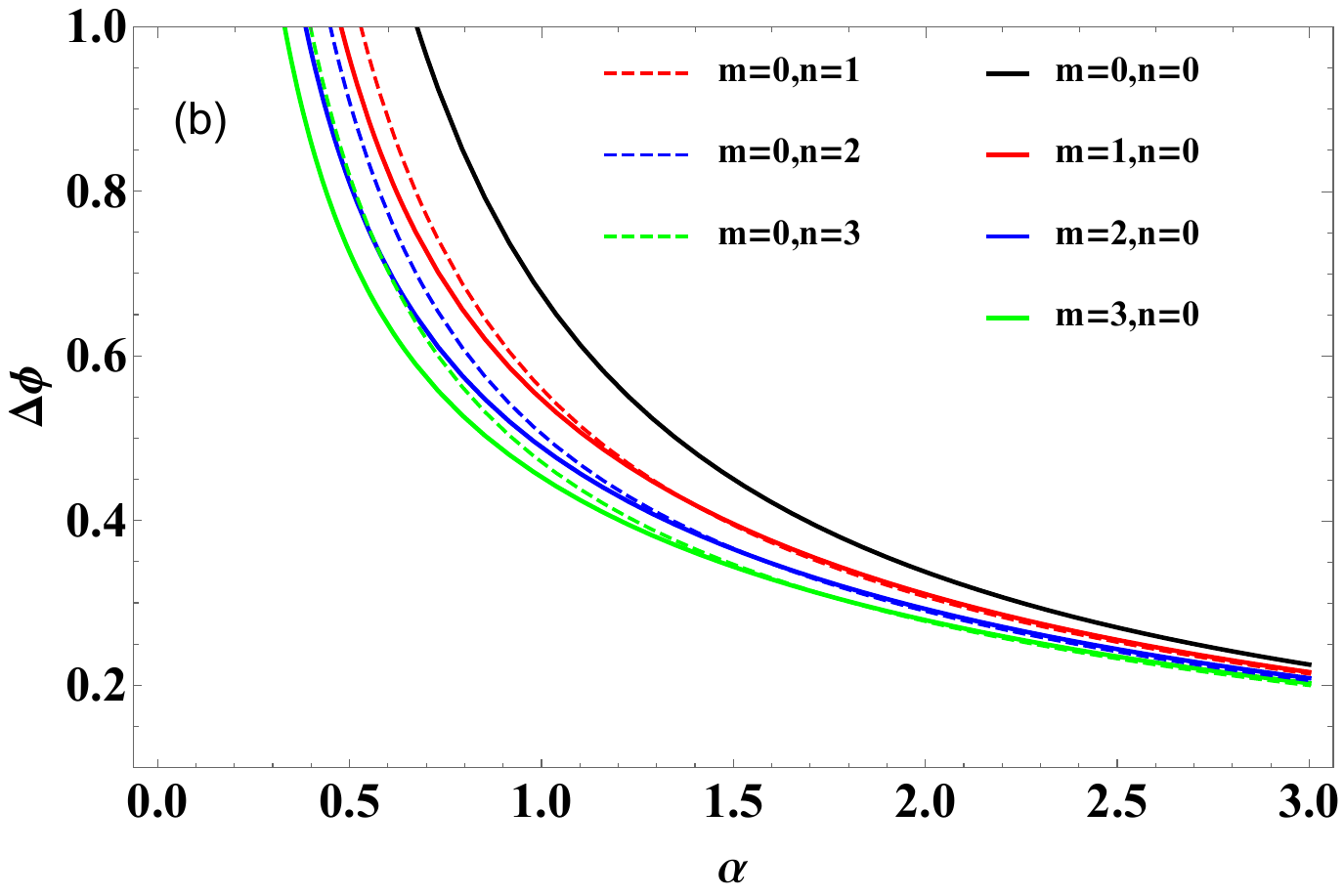}
\end{minipage}}
\caption{The phase sensitivity as a function of $\protect \alpha $, with $%
g=1\ $and $\protect \phi =0.6$. (a) symmetrical two-mode multi-PSS, (b) single mode multi-PSS.}
\end{figure}

\subsection{Photon losses case}

The SU(1,1) interferometer plays a critical role in achieving high-precision
measurements. However, precision is significantly affected by photon losses,
especially internal losses. Now, we pay our attention to the internal photon
losses, corresponding to $T_{k}\in (0,1)$. The phase sensitivity is plotted
as a function of transmittance $T_{k}$ in Fig. 5 for fixed $g$, $\alpha $, $%
\phi $, with varying numbers of subtracted photons. As anticipated, the
phase sensitivity is enhanced with increasing transmittance $T_{k}$ because
lower transmittance corresponds to higher levels of internal losses, which
weaken the performance of phase estimation. The improved effects of phase
sensitivity are also evident with an increase in the number of subtracted
photons. Furthermore, it is noteworthy that the multi-PSS on mode $b$ yields
higher phase sensitivity than that on mode $a$ in a highly dissipative
region (approximately $>70\%$), while the reverse is true in a low
dissipative region (approximately $<30\%$). However, the sensitivity
difference is not significant for the latter case. This suggests that the
single-mode multi-PSS on mode $b$ exhibits more robustness than that on mode
$a$ against large internal photon losses.
\begin{figure}[tbh]
\label{Figure5} \centering%
\subfigure{
\begin{minipage}[b]{0.5\textwidth}
\includegraphics[width=0.83\textwidth]{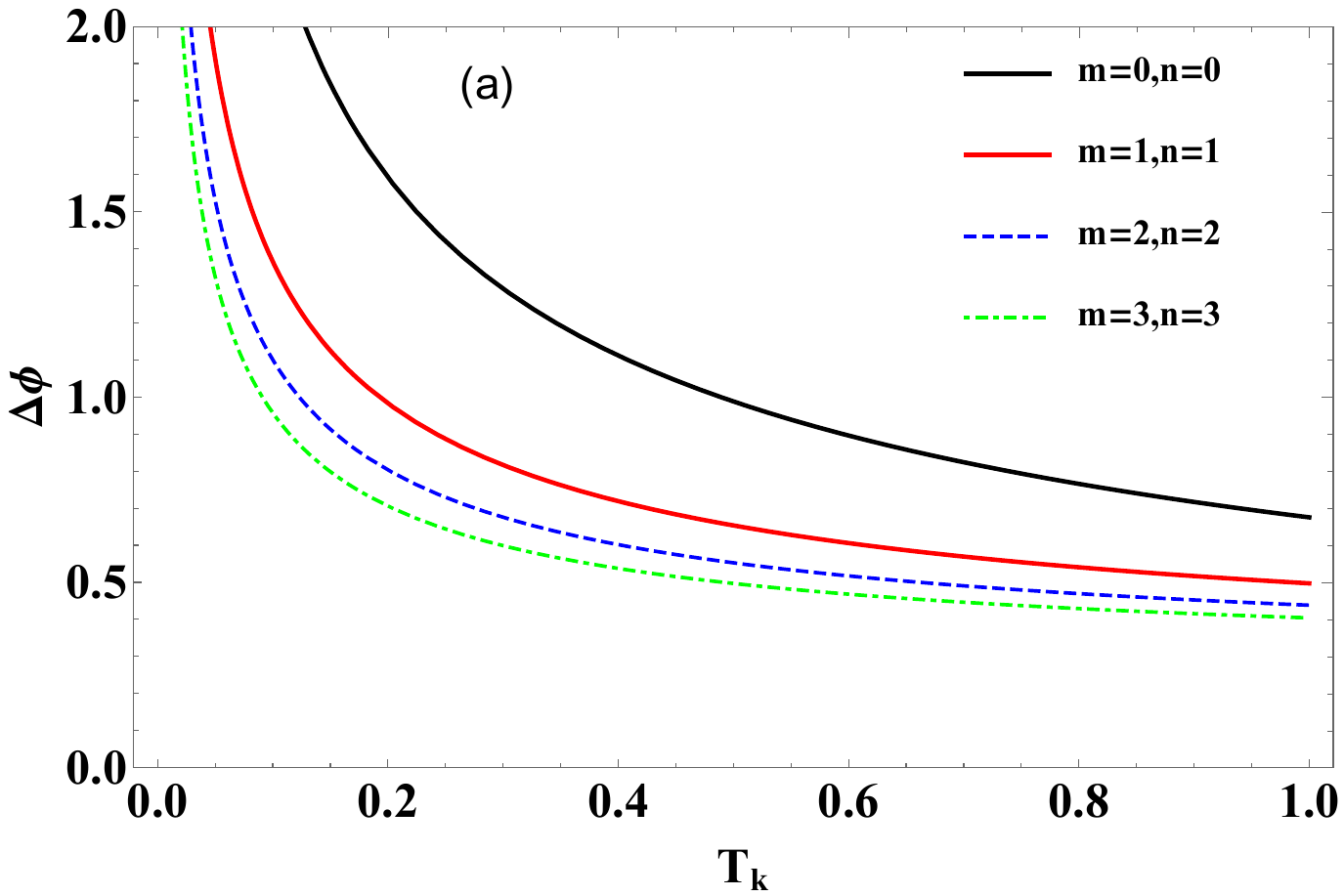}\\
\includegraphics[width=0.83\textwidth]{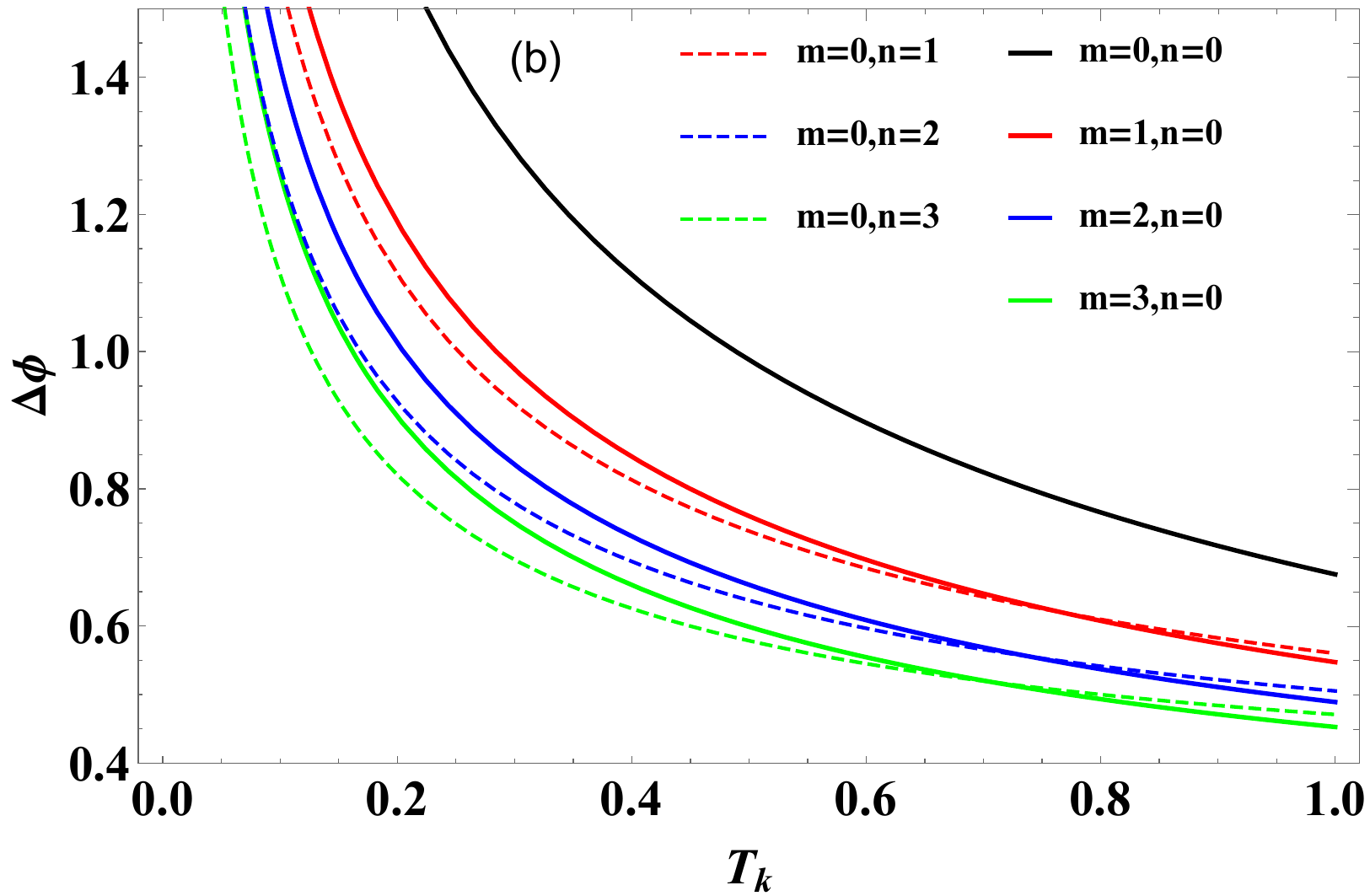}
\end{minipage}}
\caption{The phase sensitivity as a function of transmittance $T_{k}$, with $%
g=1$, $\protect \phi =0.6$ and $\protect \alpha =1.$ (a) symmetrical two-mode
multi-PSS, (b) single mode multi-PSS.}
\end{figure}

\subsection{Comparison with SQL}

Additionally, we compare the phase sensitivity with the SQL and the HL in
this subsection. The SQL and HL are defined as $\Delta \phi _{SQL}=1/\sqrt{N}
$ and $\Delta \phi _{HL}=1/N$, respectively, where $N$ represents the total
mean photon number inside the interferometer before the second OPA for each
scheme \cite{d1,d2,d3}. $N$ can be calculated as%
\begin{eqnarray}
N &=&A^{2}\langle \Psi _{in}|U_{S_{1}}^{\dagger }U_{B}^{\dagger
}U_{P}^{\dagger }\left( a^{\dagger }a+b^{\dagger }b\right)  \notag \\
&&\times U_{P}U_{B}U_{S_{1}}|\Psi _{in}\rangle  \notag \\
&=&A^{2}[D_{m_{1}+1,n_{1,}m_{2}+1,n_{2}}e^{w_{1}}  \notag \\
&&+D_{m_{1},n_{1}+1_{,}m_{2},n_{2}+1}e^{w_{1}}].  \label{a11}
\end{eqnarray}%

In our schemes, we set the total mean photon number $N=4$ for all
interferometers and compared the phase sensitivity $\Delta \phi $ with the
SQL and the HL, as depicted in Fig. 6. Our findings demonstrate that (i) the
original state (without multi-PSS) cannot surpass the SQL. (ii) Within a
wide range, the multi-PSS is capable of surpassing the SQL even in the
presence of significant photon losses (Fig. 6(b) and (c)). Additionally, the
multi-PSS can successfully surpass the SQL even for relatively high internal
losses ($T_{k}=0.5$). This suggests that the multi-PSS exhibits better
robustness against internal photon losses.
\begin{figure}[tph]
\label{Figure6} \centering
\subfigure{
\begin{minipage}[b]{0.5\textwidth}
\includegraphics[width=0.83\textwidth]{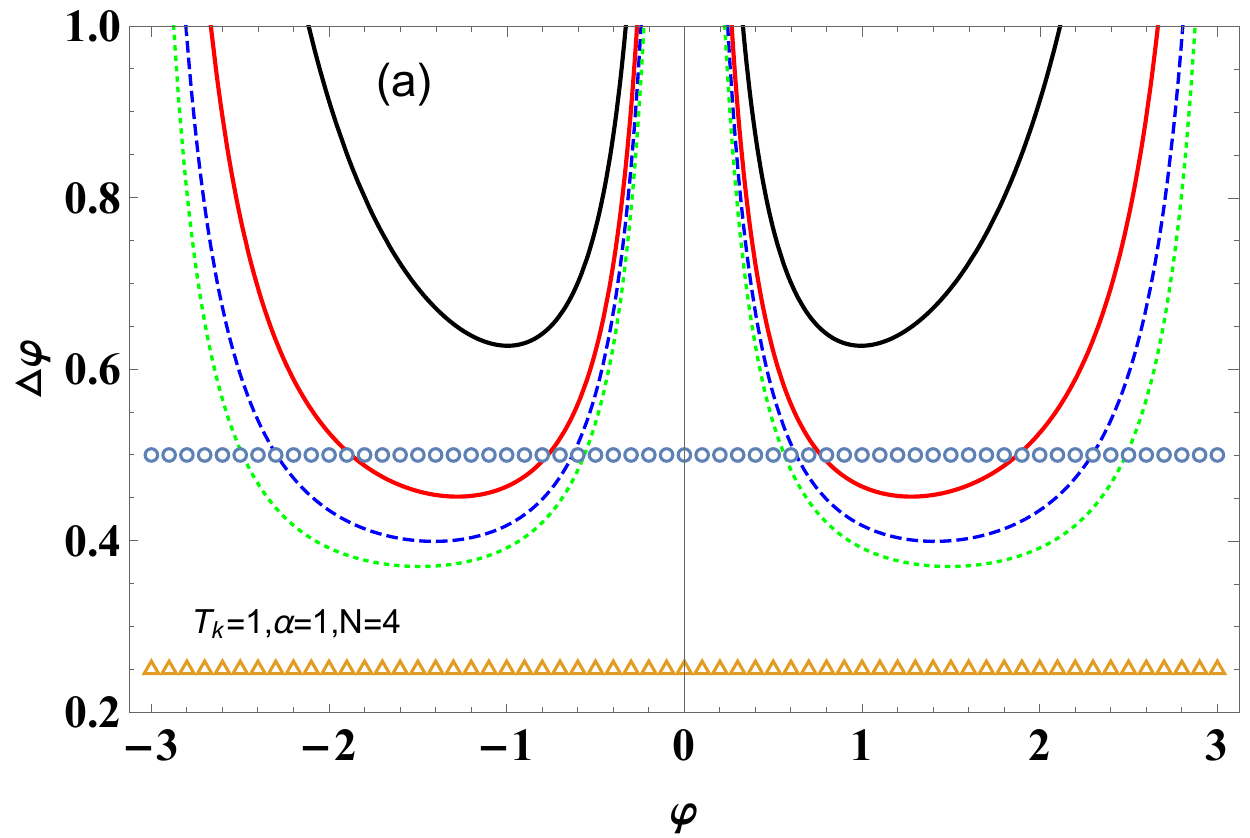}\\
\includegraphics[width=0.83\textwidth]{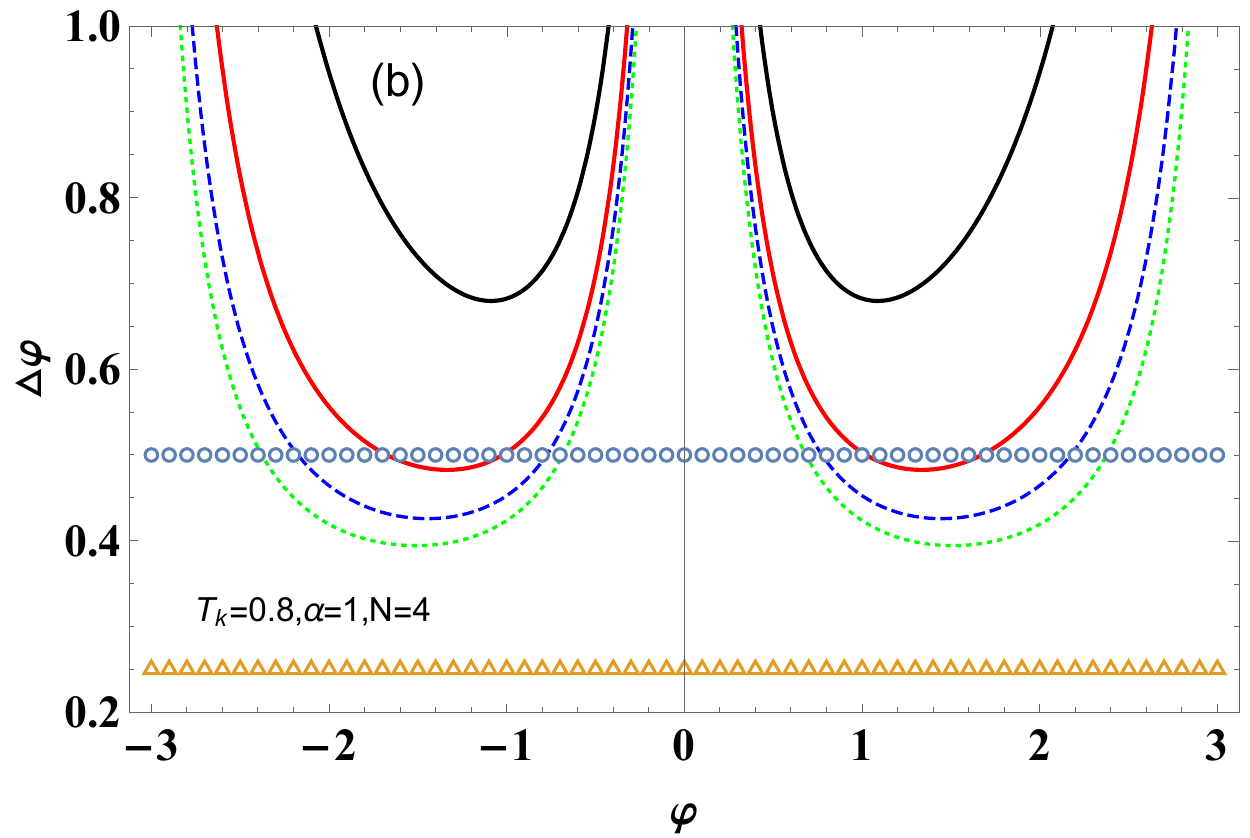}\\
\includegraphics[width=0.83\textwidth]{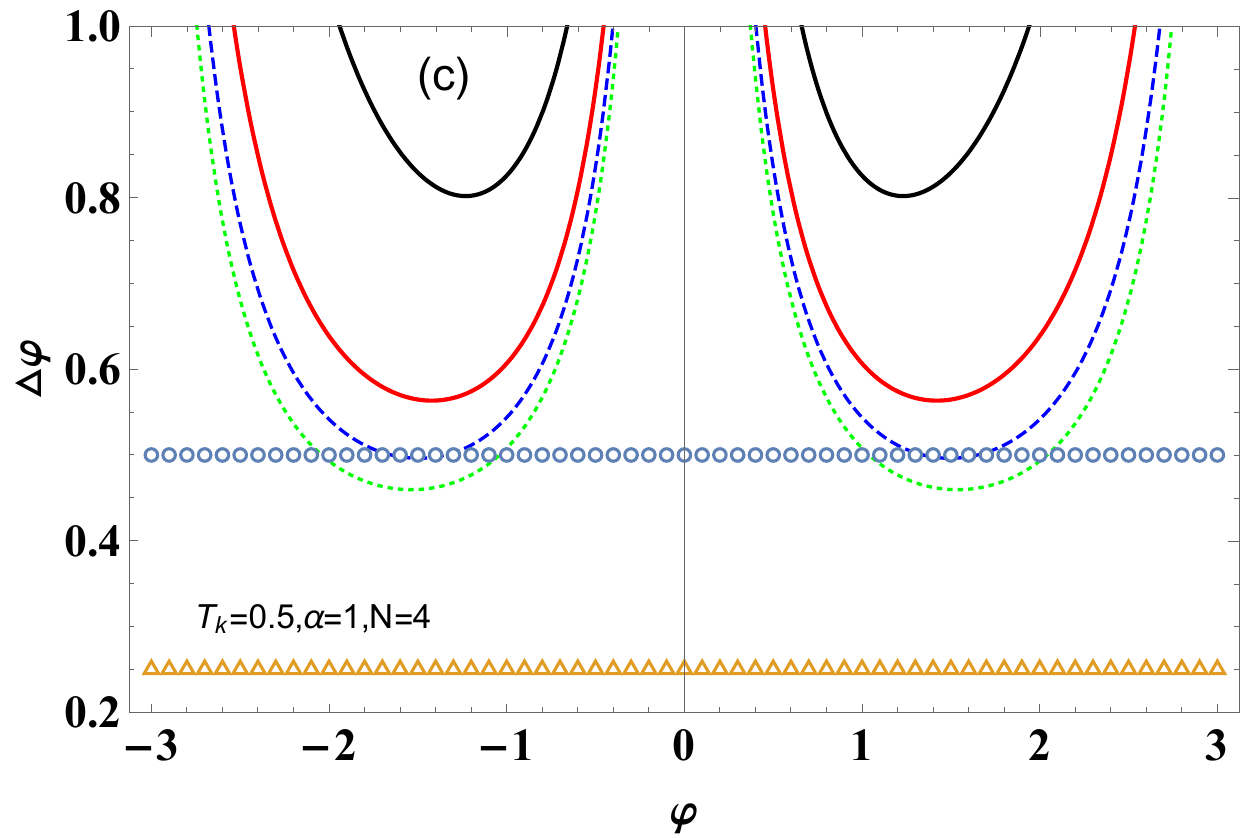}
\end{minipage}}
\caption{Comparison of the phase sensitivity with SQL and HL. In the figure,
the blue circle is SQL and the yellow triangle is HL. The black solid line
corresponds to the standard SU (1,1) interferometer, the red solid line, the
blue dashed line and the green dotted line correspond to the simultaneous
deduction of one photon, two photons and three photons from the dual modes,
respectively.}
\end{figure}

\section{The quantum Fisher information}

In the preceding discussions, we have examined the impact of multi-PSS on
phase sensitivity and the correlation between phase sensitivity and relevant
parameters based on homodyne detection. It is important to note that the
phase sensitivity discussed above is contingent on the chosen measurement
method. Then, for a given interferometer, how to get the maximum phase
sensitivity which does not depend on the specific measurement methods? In
this section, we turn our attention to the QFI, which represents the maximum
information acquired from the interferometer system, regardless of the
specific measurement method. We will cases the QFI under ideal and realistic
cases, respectively.

\subsection{Ideal case}

For a pure state system, the QFI can be derived by \cite{b12}%
\begin{equation}
F=4\left[ \left \langle \Psi _{\phi }^{\prime }|\Psi _{\phi }^{\prime
}\right \rangle -\left \vert \left \langle \Psi _{\phi }^{\prime }|\Psi
_{\phi }\right \rangle \right \vert ^{2}\right] ,  \label{a12}
\end{equation}%
where\ $\left \vert \Psi _{\phi }\right \rangle $ is the quantum state after
phase shift and before the second OPA, and $\left \vert \Psi _{\phi
}^{\prime }\right \rangle =\partial \left \vert \Psi _{\phi }\right \rangle
/\partial \phi .$ Then the QFI can be reformed as \cite{b12}
\begin{equation}
F=4\left \langle \Delta ^{2}n_{a}\right \rangle ,  \label{a13}
\end{equation}%
where $\left \langle \Delta ^{2}n_{a}\right \rangle =\left \langle \Psi
_{\phi }\right \vert (a^{\dagger }a)^{2}|\Psi _{\phi }\rangle -(\left
\langle \Psi _{\phi }\right \vert a^{\dagger }a|\Psi _{\phi }\rangle )^{2}$.

In the ideal multi-PSS, the quantum state is given by $\left \vert \Psi
_{\phi }\right \rangle =AU_{\phi }U_{p}U_{S_{1}}|\Psi _{in}\rangle $\textit{%
\ }with\textit{\ }$|\Psi _{in}\rangle =\left \vert \alpha \right \rangle
_{a}\otimes \left \vert 0\right \rangle _{b}$, and $U_{p}=a^{m}\otimes b^{n}.
$ Thus, the QFI is derived as
\begin{eqnarray}
F &=&4[A^{2}D_{m_{1}+2,n_{1,}m_{2}+2,n_{2}}e^{w_{1}}  \notag \\
&&+A^{2}D_{m_{1}+1,n_{1,}m_{2}+1,n_{2}}e^{w_{1}}  \notag \\
&&-\left( A^{2}D_{m_{1}+1,n_{1,}m_{2}+1,n_{2}}e^{w_{1}}\right) ^{2}].
\label{a14}
\end{eqnarray}%
In the above equations, $T_{k}=1$. It is possible to explore the connection
between the QFI and the related parameters using Eqs. (\ref{a14}).

Fig. 7 and Fig. 8 depict the QFI as a function of $g$ ($\alpha $) for a
specific $\alpha $ ($g$). It is evident that a higher value of $g$ ($\alpha $%
) corresponds to greater QFI. Additionally, it is observed that the QFI is
enhanced for the multi-PSS due to the non-Gaussian operation, and this
enhancement further increases with the number of manipulated photons.
Furthermore, the multi-PSS exhibits slightly better performance on mode $b$
than on mode $a$ (Fig. 7(b) and 8(b)).
\begin{figure}[tbh]
\label{Figure7} \centering%
\subfigure{
\begin{minipage}[b]{0.5\textwidth}
\includegraphics[width=0.83\textwidth]{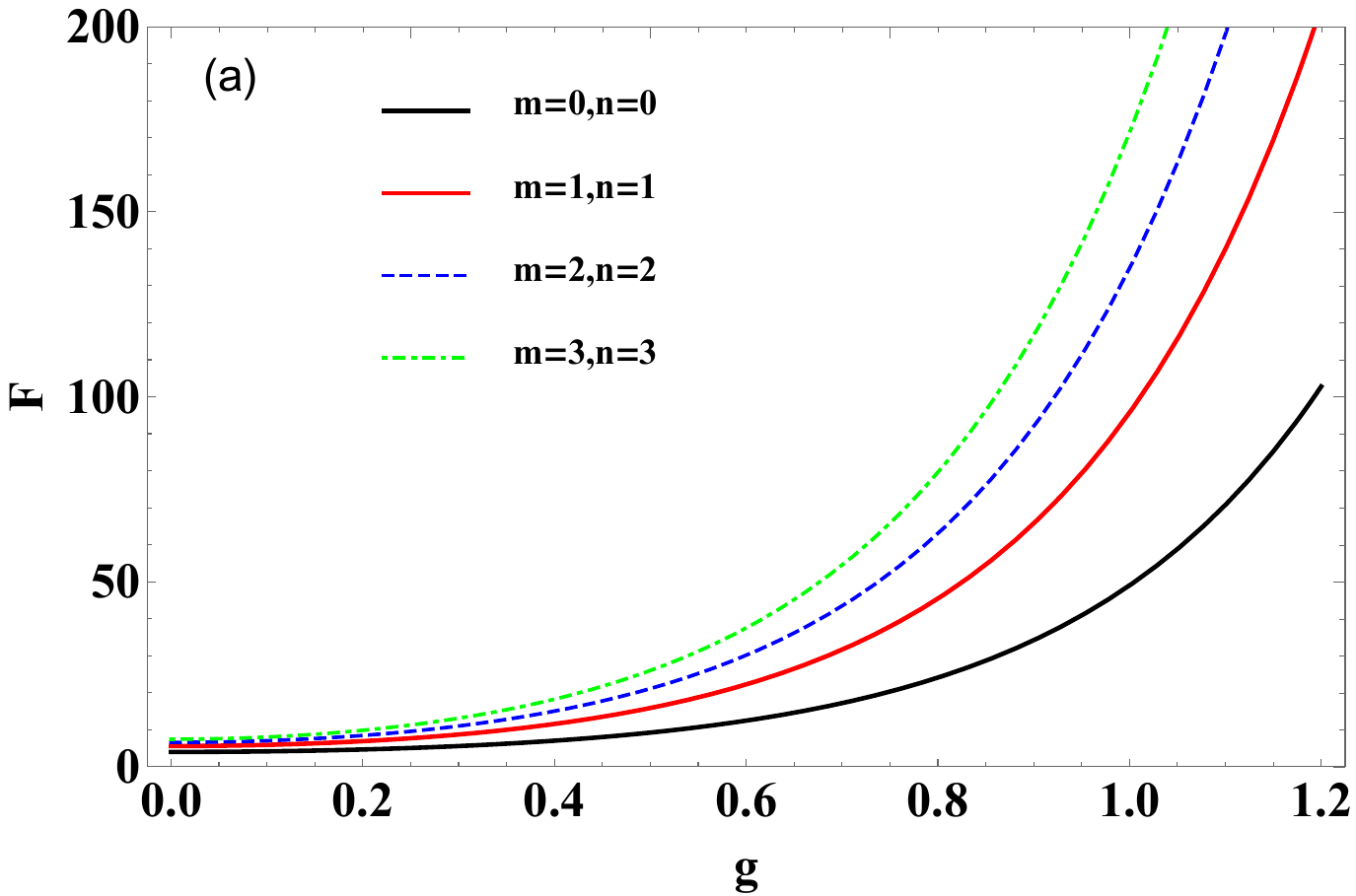}\\
\includegraphics[width=0.83\textwidth]{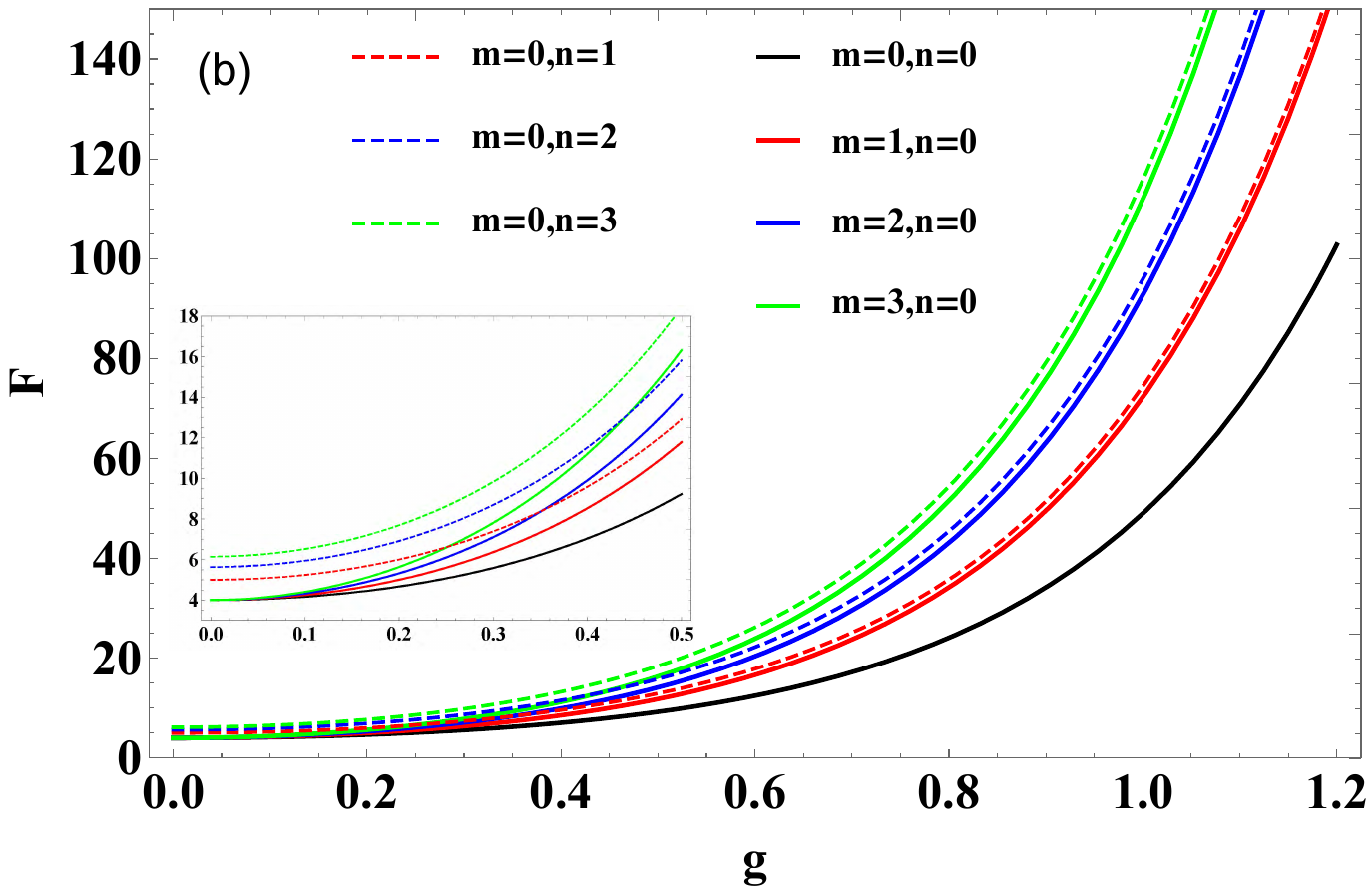}
\end{minipage}}
\caption{The QFI as a function of $g$, with $\protect \alpha =1$.
(a) symmetrical two-mode multi-PSS, (b) single mode multi-PSS.}
\end{figure}
\begin{figure}[tbh]
\label{Figure8} \centering%
\subfigure{
\begin{minipage}[b]{0.5\textwidth}
\includegraphics[width=0.83\textwidth]{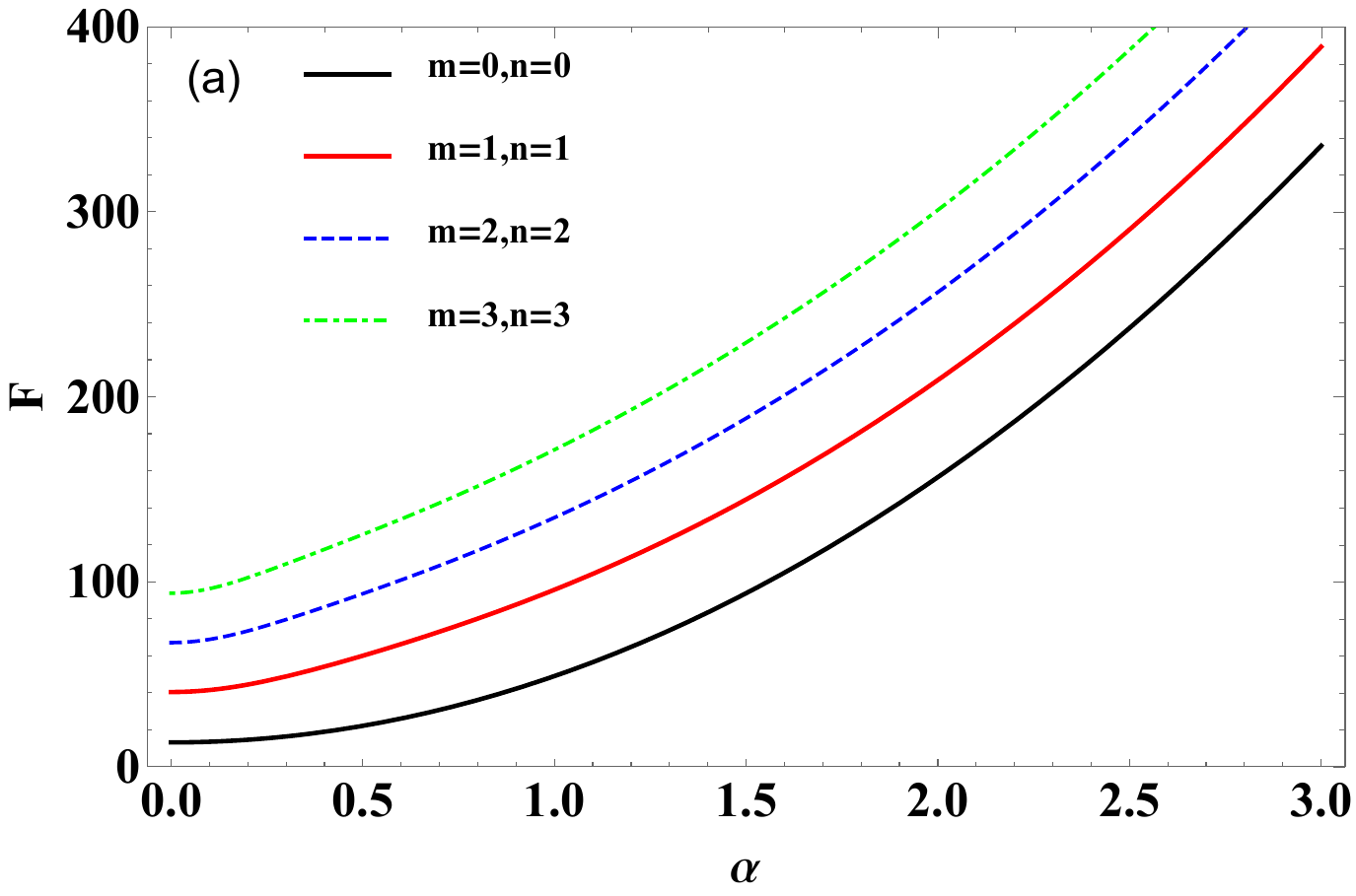}\\
\includegraphics[width=0.83\textwidth]{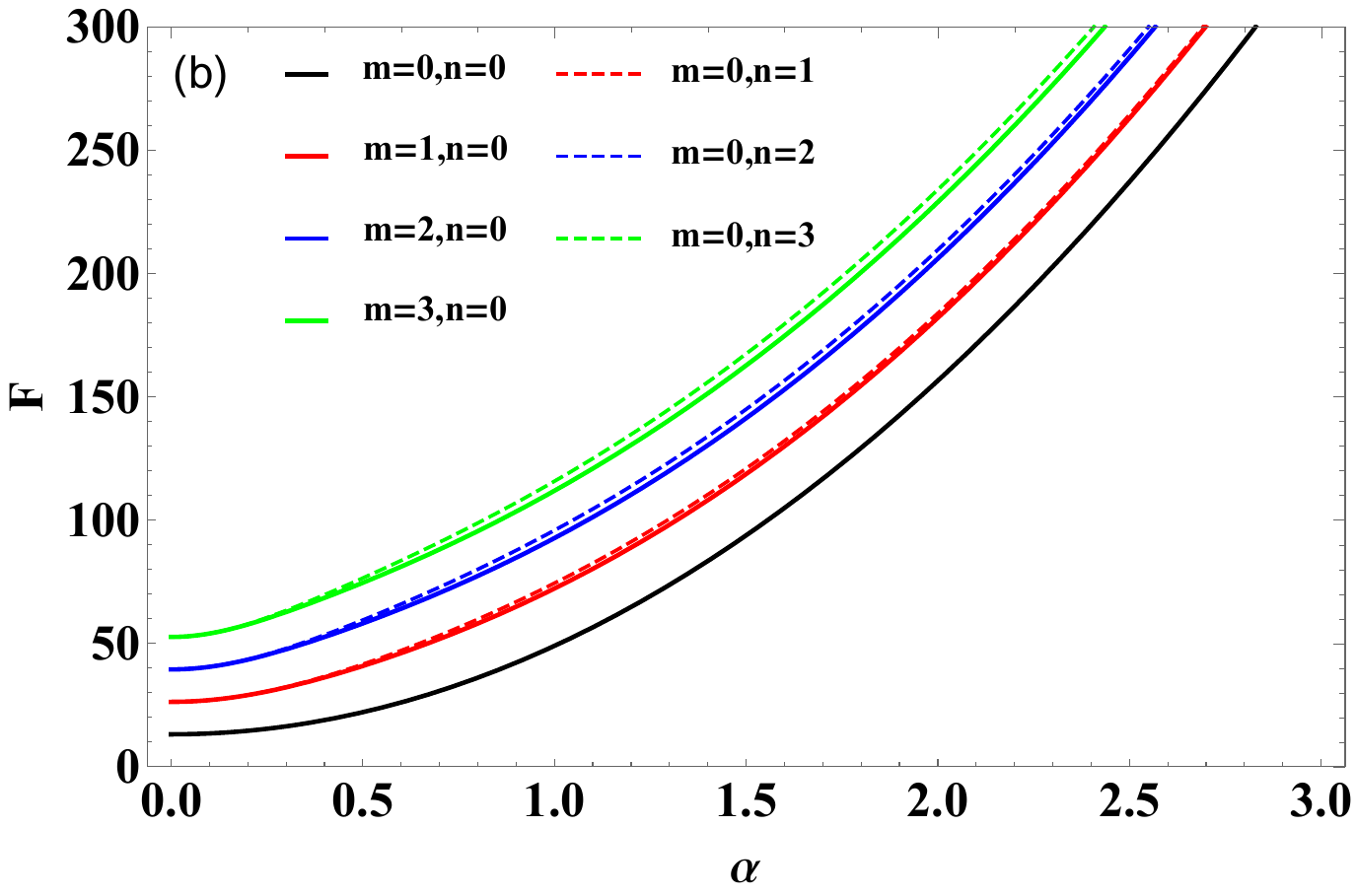}
\end{minipage}}
\caption{The QFI as a function of $\protect \alpha $, with $g=1$.
(a) symmetrical two-mode multi-PSS, (b) single mode multi-PSS.}
\end{figure}

Actually, the QFI can be related with the phase sensitivity via \cite{b13}%
\begin{equation}
\Delta \phi _{QCRB}=\frac{1}{\sqrt{vF}},  \label{a15}
\end{equation}%
Where $v$ is the number of measurements. For simplicity, we set $v=1$.
Another quantum limit, the QCRB \cite{b3,b4}, is denoted as $\Delta \phi
_{QCRB}$, and it establishes the ultimate limit for a set of probabilities
resulting from measurements on a quantum system. It is an estimator
implemented asymptotically by a maximum likelihood estimator and provides a
detection-independent phase sensitivity.

Fig. 9 and Fig. 10 illustrate the variation of $\Delta \phi _{QCRB}$ as a
function of $g$ ($\alpha $) for a specific $\alpha $ ($g$). It is shown that
$\Delta \phi _{QCRB}$ improves with increasing $g$ and $\alpha $.
Additionally, as the number of photon operations increases, the multi-PSS
exhibits greater enhancement for $\Delta \phi _{QCRB}$. Overall, the
multi-PSS exhibits better performance on mode $b$ compared to mode $a$,
especially when the gain factor $g$ is smaller (refer to Fig. 9(b)).
Furthermore, the improvement in $\Delta \phi _{QCRB}$ is more obvious for
small coherent amplitude $\alpha $ (refer to Fig. 10).
\begin{figure}[tbh]
\label{Figure9} \centering%
\subfigure{
\begin{minipage}[b]{0.5\textwidth}
\includegraphics[width=0.83\textwidth]{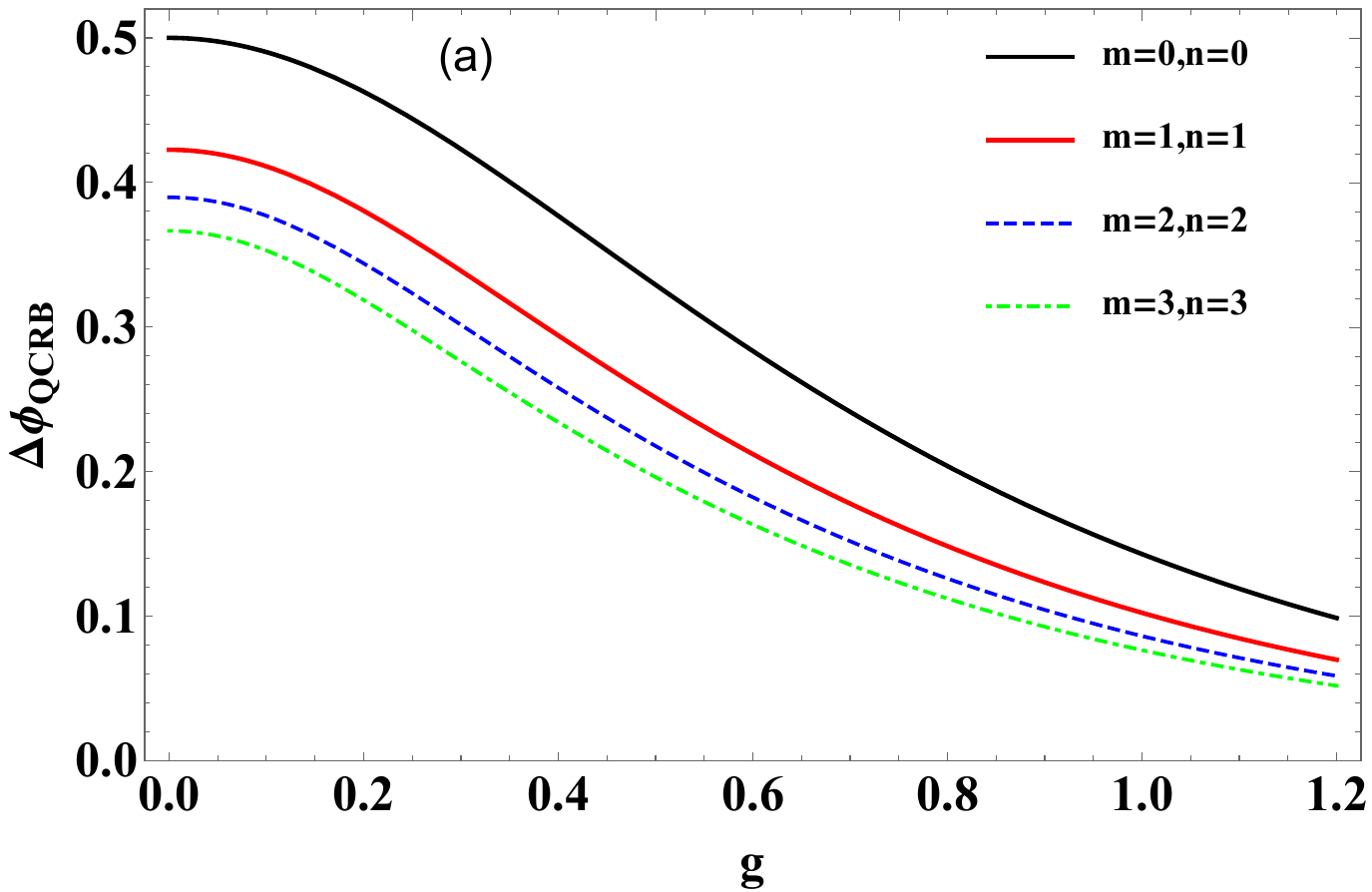}\\
\includegraphics[width=0.83\textwidth]{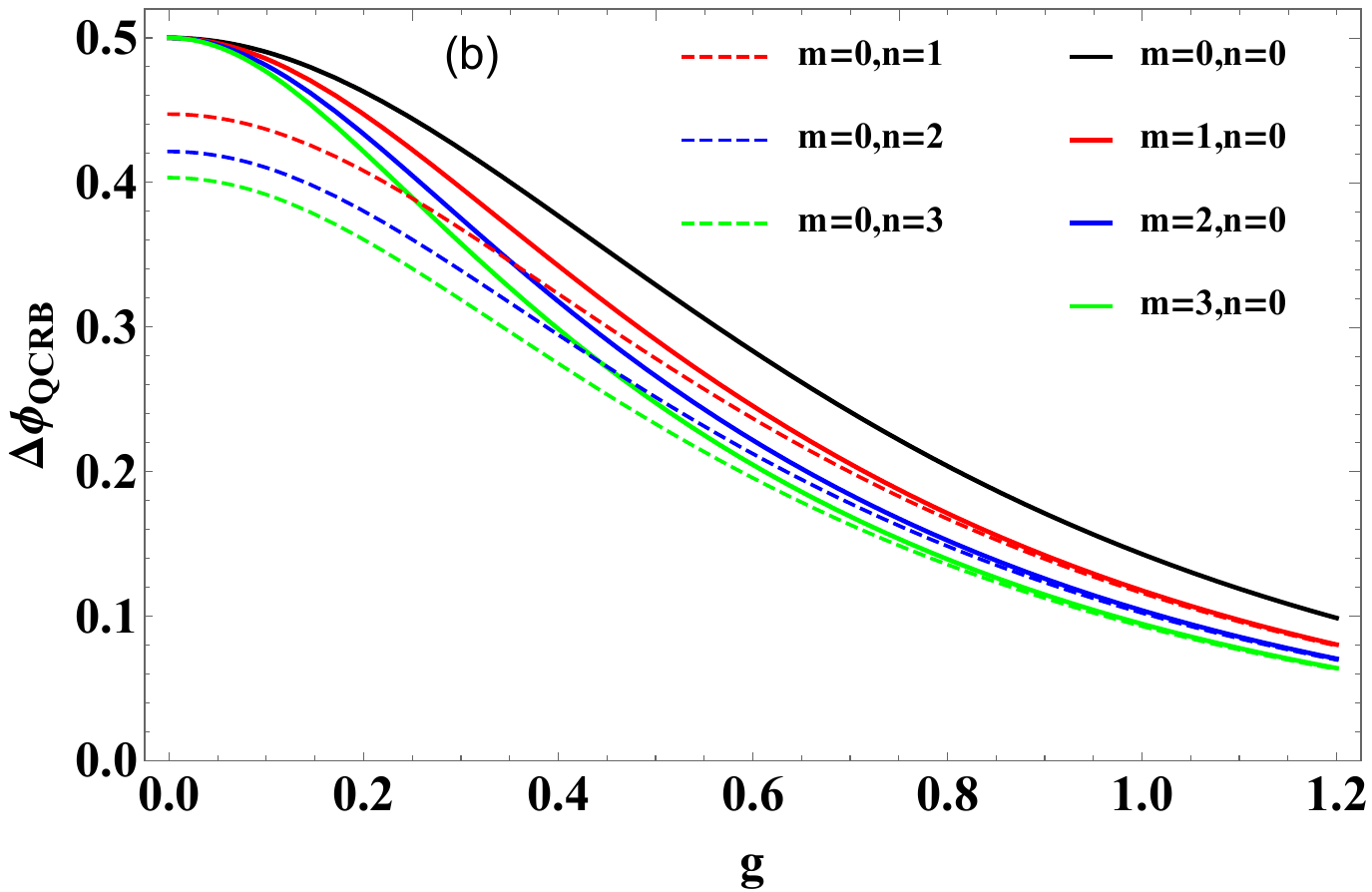}
\end{minipage}}
\caption{The $\Delta \protect \phi _{QCRB}$ as a function of $g$, with $%
\protect \alpha =1$. (a) symmetrical two-mode multi-PSS, (b) single mode multi-PSS.}
\end{figure}
\begin{figure}[tbh]
\label{Figure10} \centering%
\subfigure{
\begin{minipage}[b]{0.5\textwidth}
\includegraphics[width=0.83\textwidth]{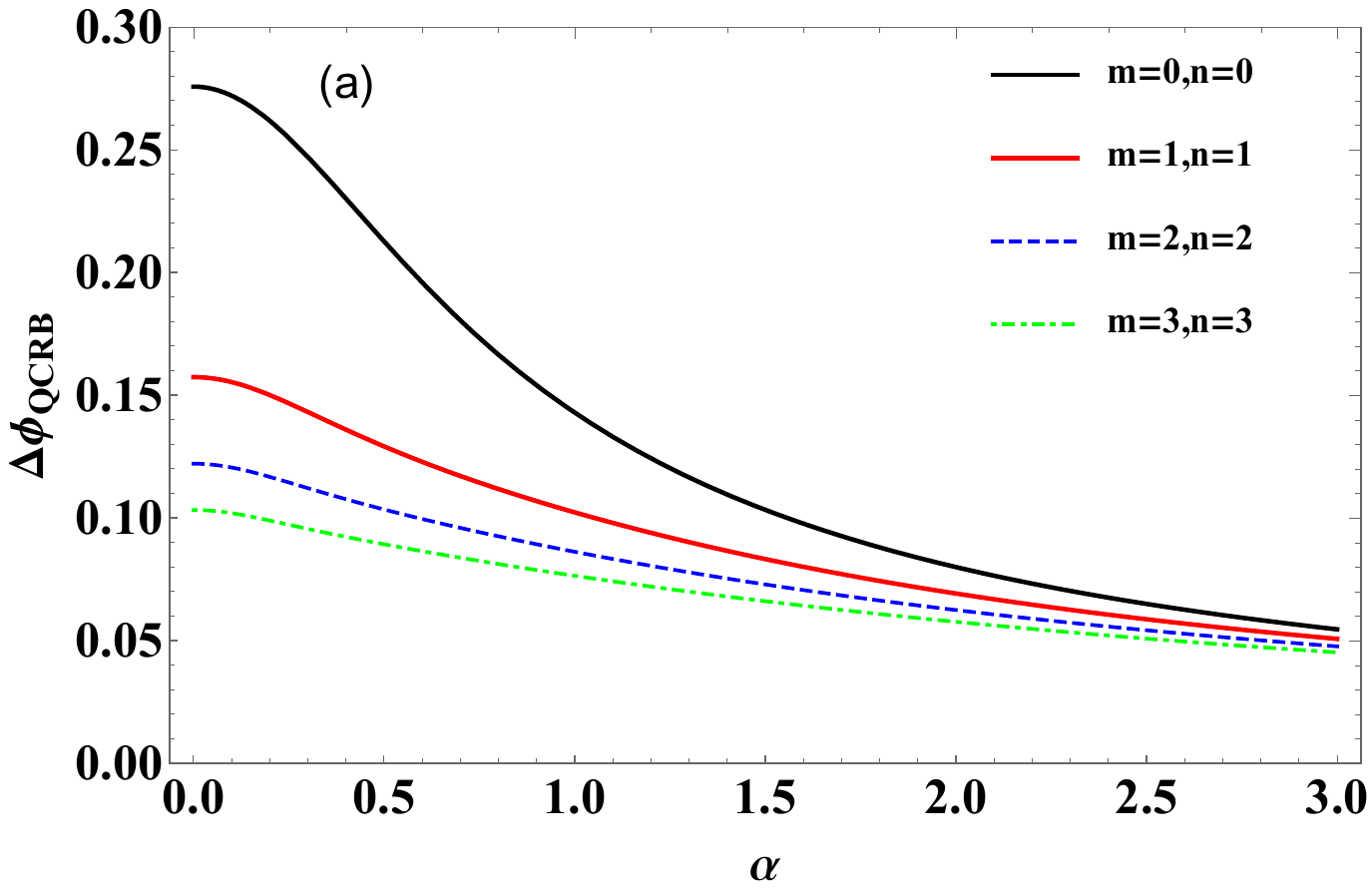}\\
\includegraphics[width=0.83\textwidth]{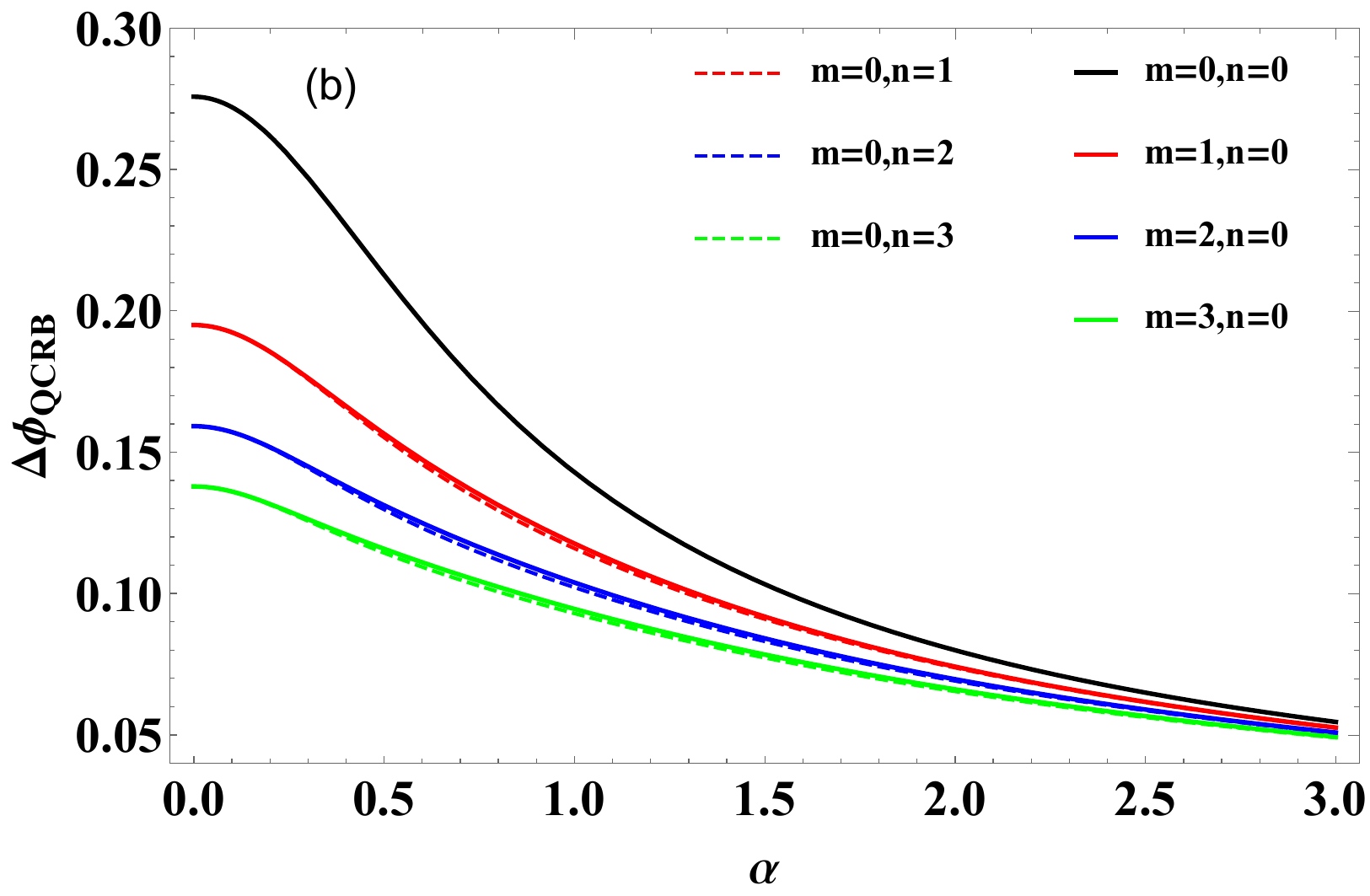}
\end{minipage}}
\caption{The $\Delta \protect \phi _{QCRB}$ as a function of $\protect \alpha $%
, with $g=1$. (a) symmetrical two-mode multi-PSS, (b) single mode multi-PSS.}
\end{figure}

In order to acquire a more comprehensive understanding of the impact of the
multi-PSS on QFI or $\Delta \phi _{QCRB}$, we have graphed the total mean
internal photon number $N$ as a function of the gain factor $g$ and the
amplitude $\alpha $, as illustrated in Fig. 11 and Fig. 12. It is evident
that the multi-PSS can increase $N$ as the number of subtracted photons
increases. Moreover, the total mean photon number for the multi-PSS on mode $%
b$ is higher than that on mode $a$.
\begin{figure}[tbh]
\label{Figure11} \centering%
\subfigure{
\begin{minipage}[b]{0.5\textwidth}
\includegraphics[width=0.83\textwidth]{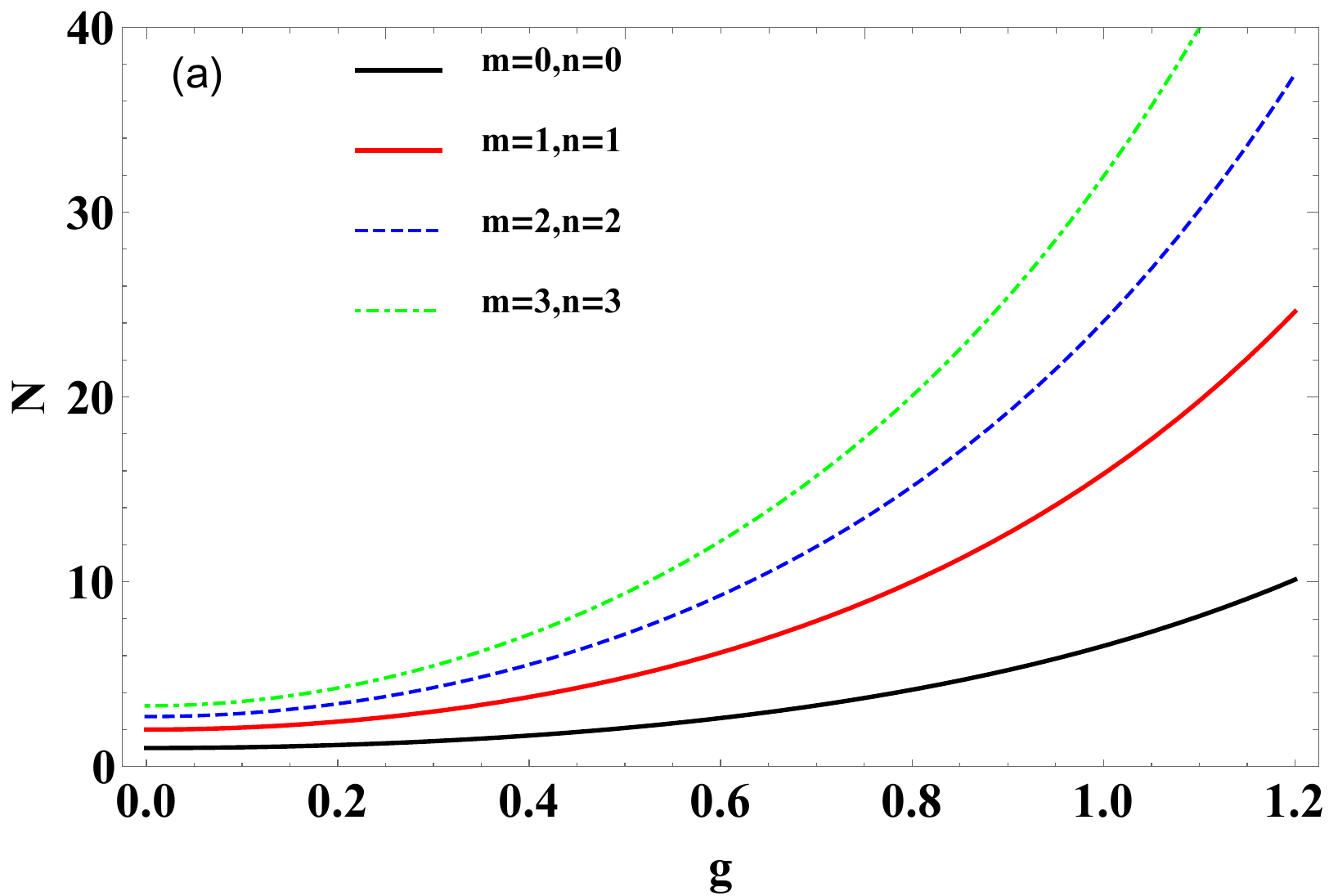}\\
\includegraphics[width=0.83\textwidth]{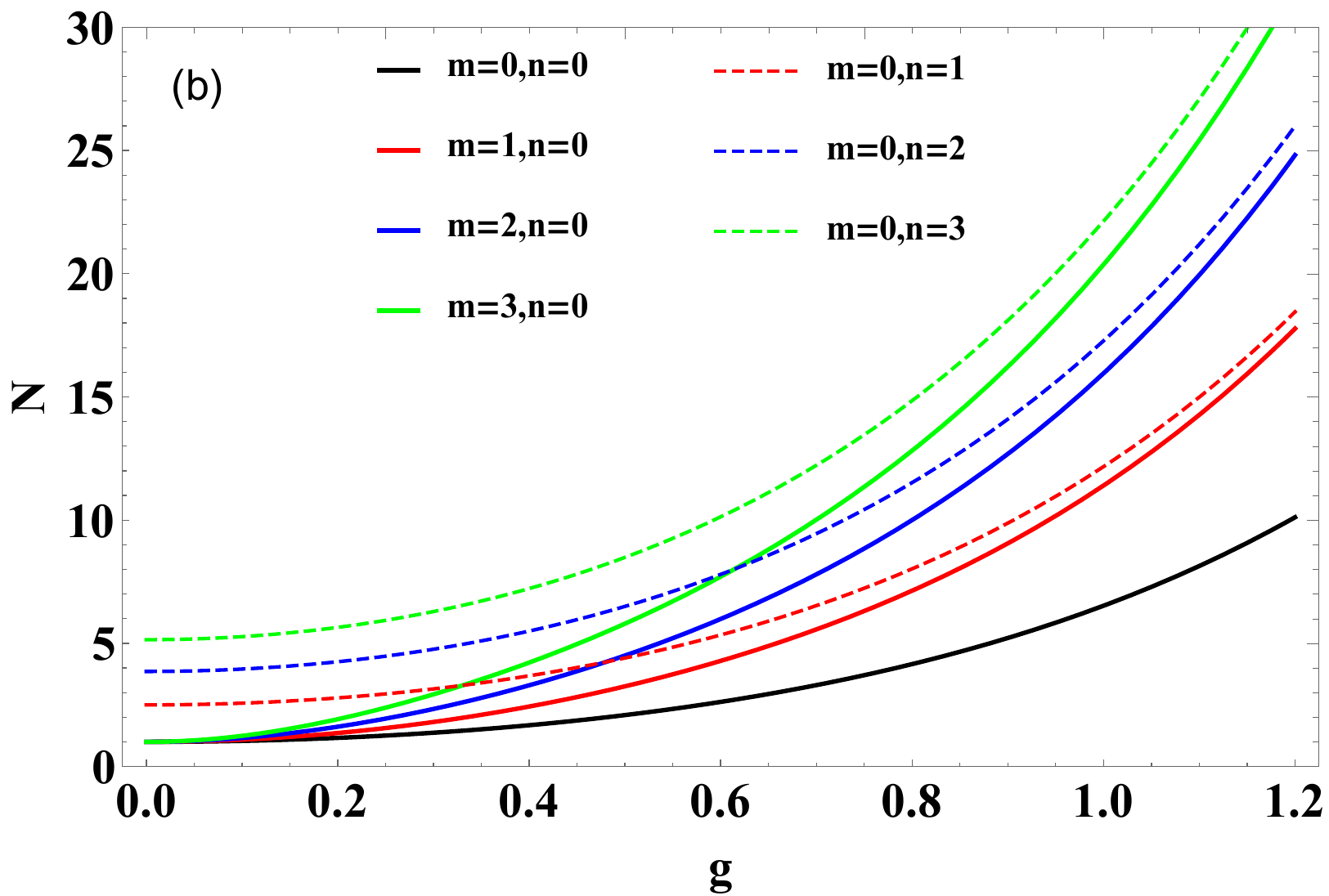}
\end{minipage}}
\caption{The total mean photon number $N$ as a function of $g$, with $%
\protect \alpha =1$. (a) symmetrical two-mode multi-PSS, (b) single mode multi-PSS.}
\end{figure}
\begin{figure}[tbh]
\label{Figure12} \centering%
\subfigure{
\begin{minipage}[b]{0.5\textwidth}
\includegraphics[width=0.83\textwidth]{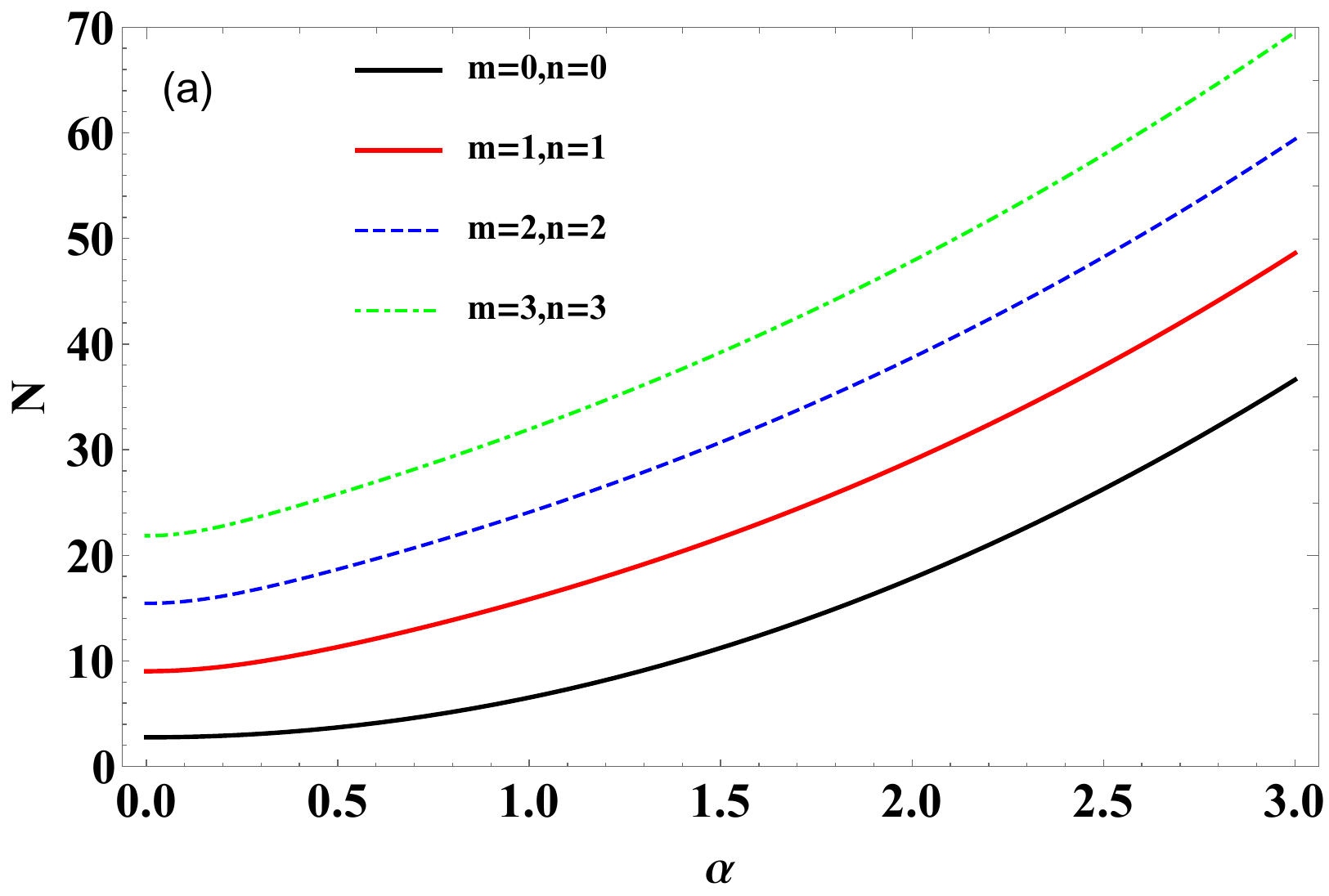}\\
\includegraphics[width=0.83\textwidth]{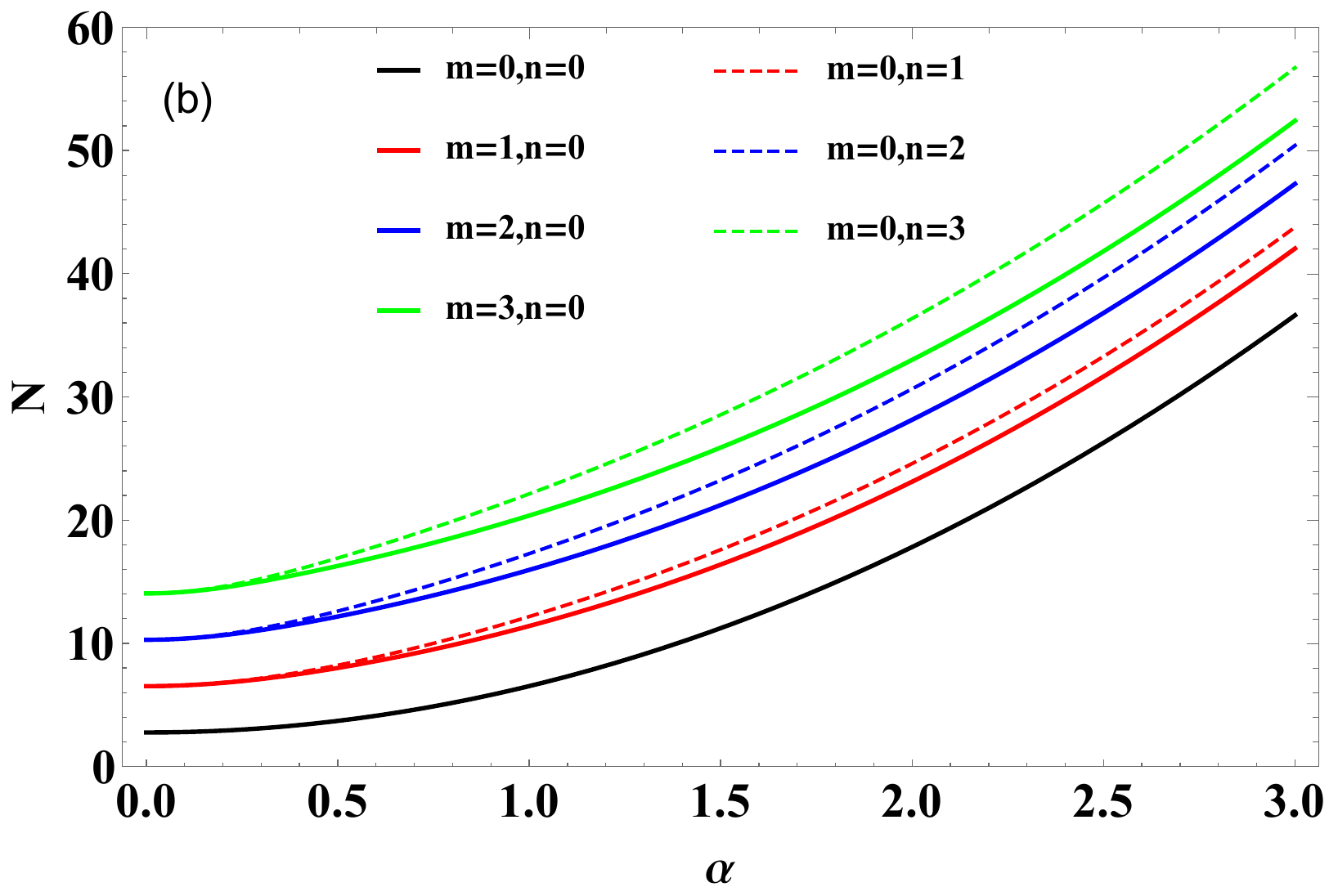}
\end{minipage}}
\caption{The total mean photon number $N$ as a function of $\protect \alpha $%
, with $g=1$. (a) symmetrical two-mode multi-PSS, (b) single mode multi-PSS.}
\end{figure}

\subsection{Photon losses case}

In this subsection, we expand our analysis to encompass the QFI
in the presence of photon losses. We consider homodyne detection on mode $a$%
, which is susceptible to photon losses on mode $a$. Consequently, our
attention is directed toward the QFI of mode $a$ under photon losses, as
illustrated in Fig. 13. For realistic quantum systems, we have verified the
feasibility of computing the QFI with internal non-Gaussian operations by
redefining the Kraus operator according to the method proposed by Escher
\emph{et al}. \cite{b12}. The detailed computational procedure is outlined
in Appendix B. Simplifying the calculation process allows us to derive the
QFI under photon losses \cite{b14}.
\begin{equation}
F_{L}=\frac{4F\eta \left \langle n_{a}\right \rangle }{\left( 1-\eta \right)
F+4\eta \left \langle n_{a}\right \rangle },  \label{a16}
\end{equation}%
where $F$ is the QFI in the ideal case, and $\eta $ is the\ transmittance
\cite{b15}. Hence, in the presence of photon losses, the QFI in our scheme
can be expressed by the following equation
\begin{equation}
F_{L}=\frac{4F\eta \left(
A^{2}D_{m_{1}+1,n_{1},m_{2}+1,n_{2}}e^{w_{1}}\right) }{\left( 1-\eta \right)
F+4\eta \left( A^{2}D_{m_{1}+1,n_{1},m_{2}+1,n_{2}}e^{w_{1}}\right) }.
\label{a17}
\end{equation}
\begin{figure}[tbh]
\label{Figure13} \centering \includegraphics[width=0.9%
\columnwidth]{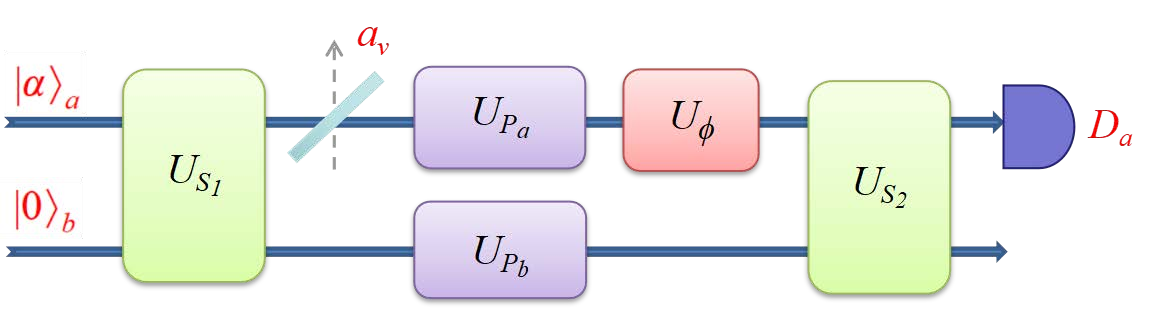}
\caption{Schematic diagram of the photon losses on mode $a$. the losses occurs before
the multi-PSS.}
\end{figure}

Under the condition of photon losses, we analyze the effects of each
parameter on the QFI to further characterize the degradation of QFI due to
photon losses. In Fig. 14, it can be observed that the QFI increases as the
transmittance $\eta $ increases, which is further enhanced as the
photon-subtracted number increases. Similar to the ideal case of QFI, this
can be attributed to the multi-PSS increasing the number of photons
internally, resulting in higher quantum information. It is noteworthy that
for both symmetrical and asymmetrical multi-PSS, the improved QFI increases
with the transmittance $\eta $. Furthermore, the multi-PSS on mode $b$
exhibits a higher QFI than that on mode $a$ (Fig. 14(b)).
\begin{figure}[tbh]
\label{Figure14} \centering%
\subfigure{
\begin{minipage}[b]{0.5\textwidth}
\includegraphics[width=0.83\textwidth]{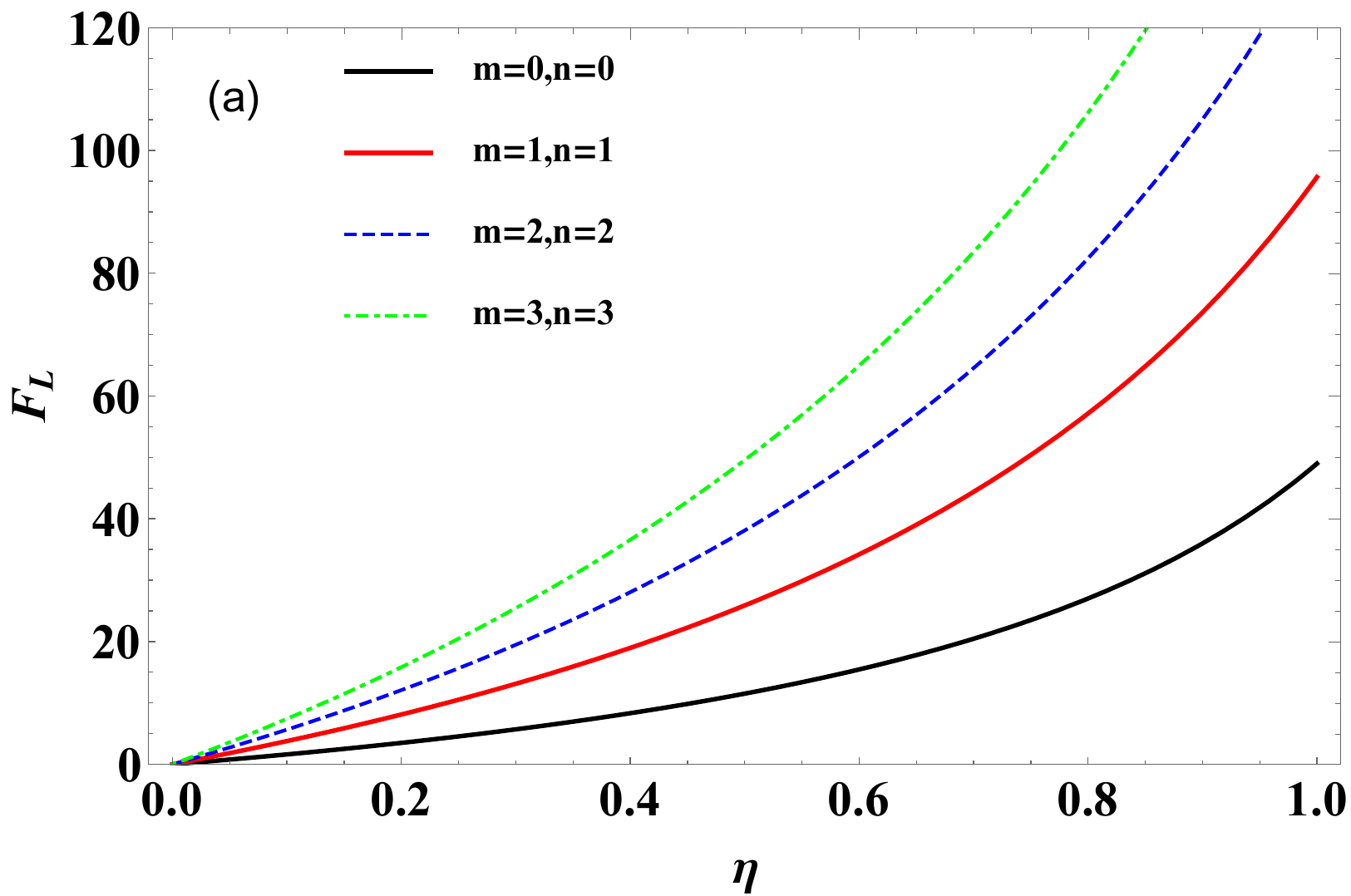}\\
\includegraphics[width=0.83\textwidth]{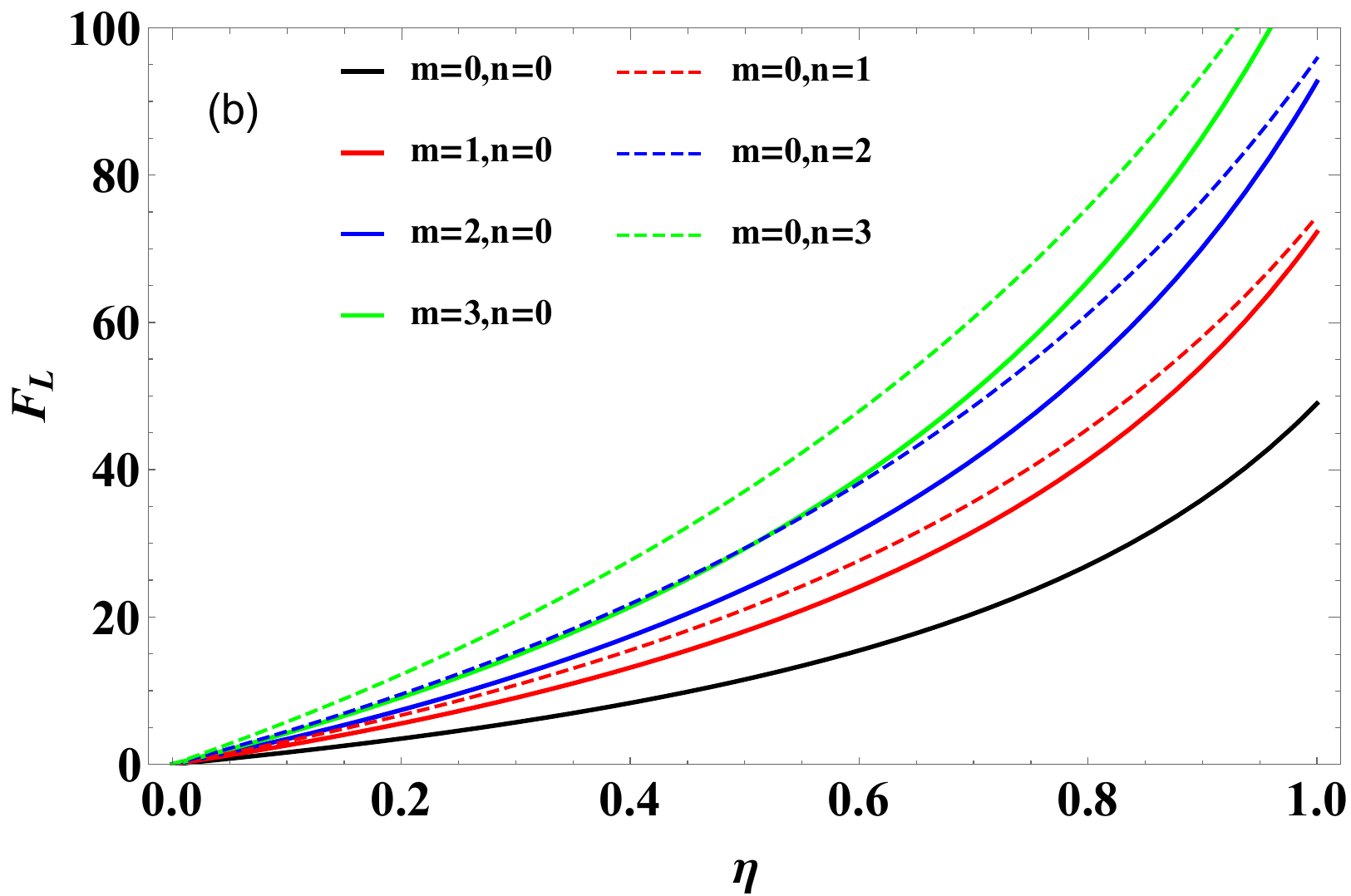}
\end{minipage}}
\caption{The $F_{L}$ as a function of transmittance $\protect \eta $, with $%
g=1$ and $\protect \alpha =1$. (a) symmetrical two-mode multi-PSS, (b) single mode multi-PSS.}
\end{figure}

Fig. 15 and Fig. 16 illustrate the QFI as a function of $g$ ($\alpha $) for
a given $\alpha $ ($g$), with $\eta =0.8$. In general, the QFI increases as $%
g$ ($\alpha $) increases. With an increase in the photon-subtracted number,
the multi-PSS exhibits a more pronounced enhancement in QFI. Once more, the
QFI of multi-PSS on mode $b$ surpasses that on mode $a$ (refer to Fig. 15(b)
and 16(b)). Additionally, for both symmetrical and asymmetrical multi-PSS,
the enhanced QFI increases with the gain factor $g$, while the changes in
the enhanced QFI with the coherent amplitude $\alpha $ are insignificant.
\begin{figure}[tbh]
\label{Figure15} \centering%
\subfigure{
\begin{minipage}[b]{0.5\textwidth}
\includegraphics[width=0.83\textwidth]{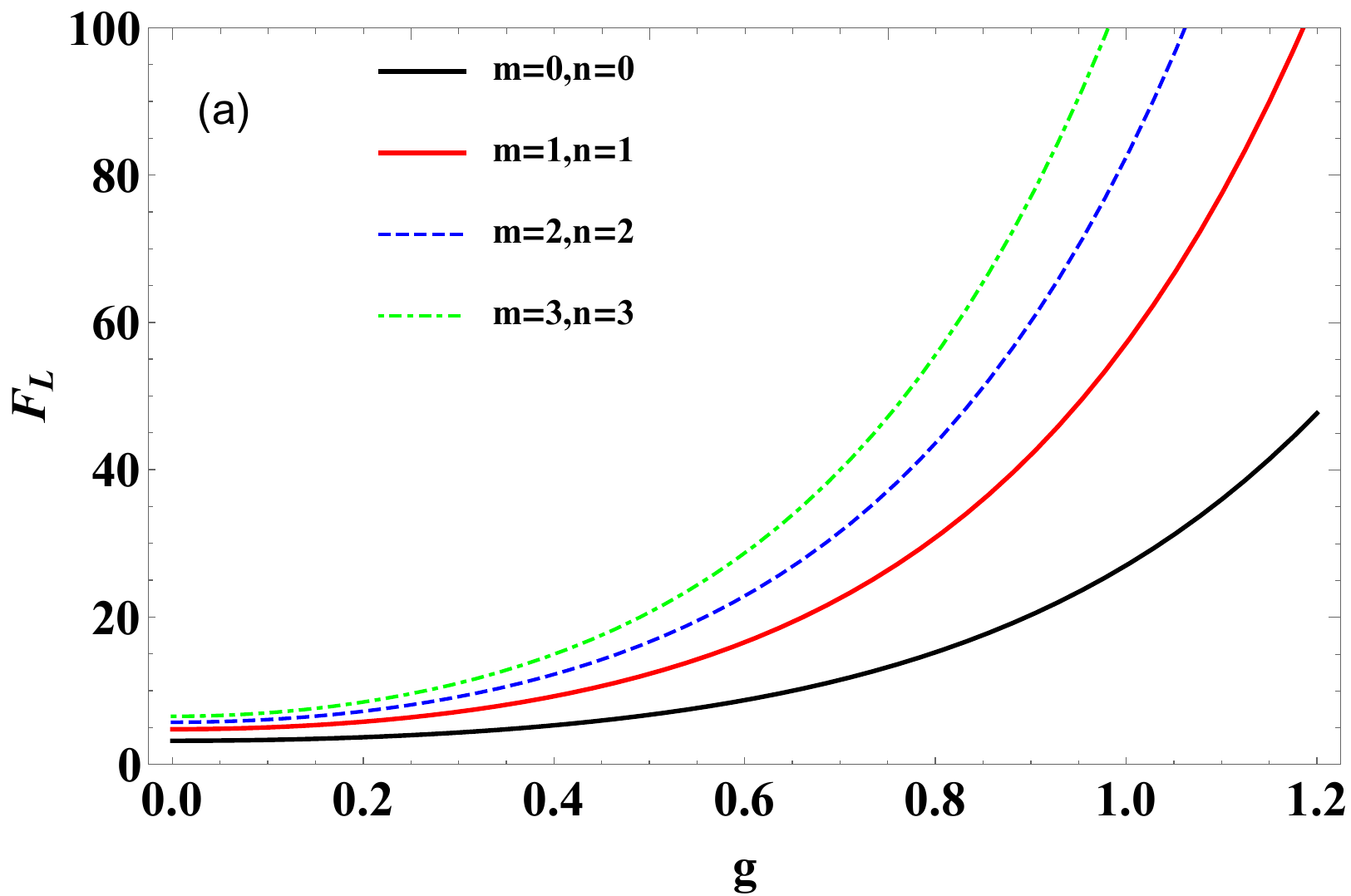}\\
\includegraphics[width=0.83\textwidth]{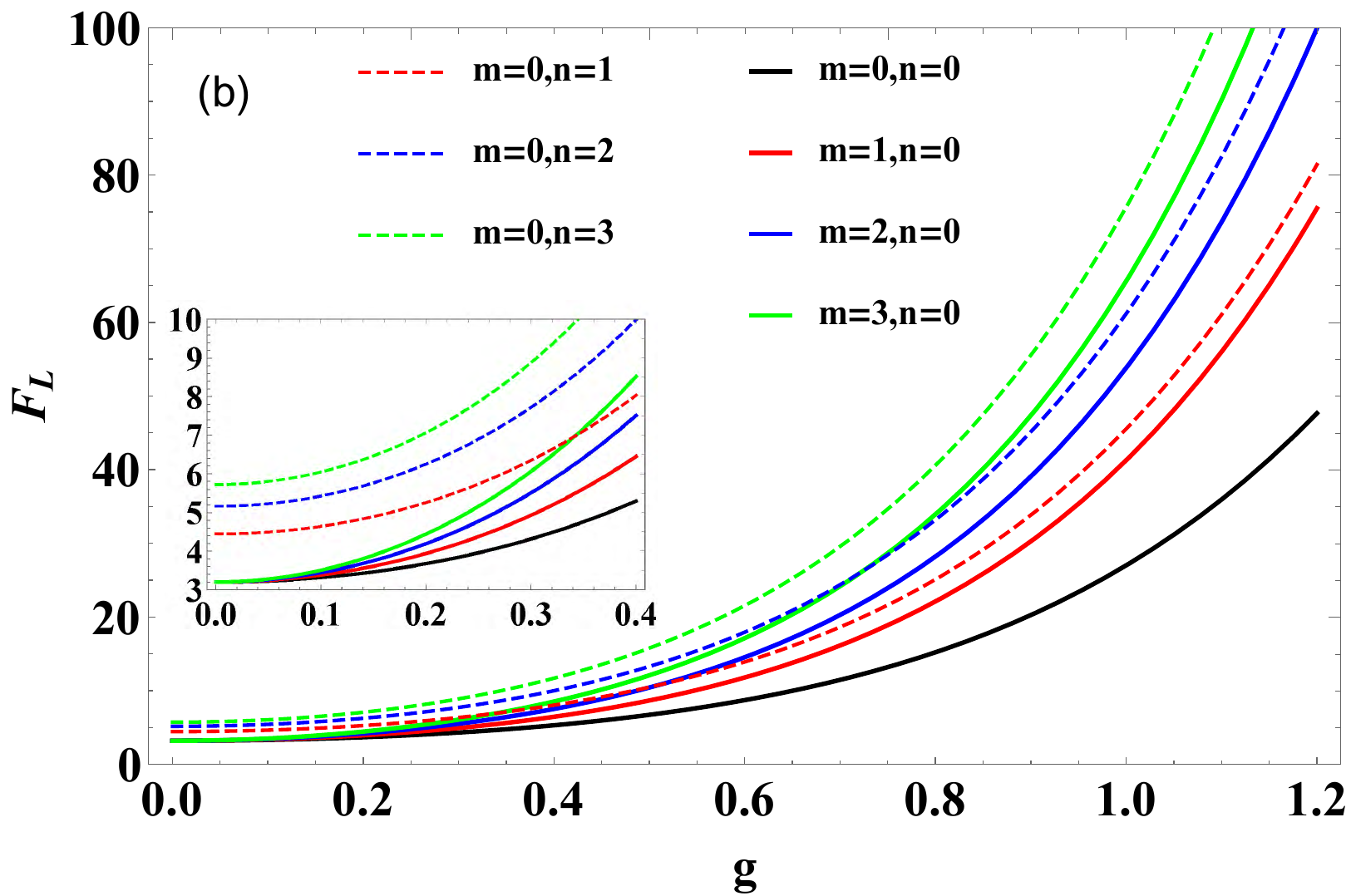}
\end{minipage}}
\caption{The $F_{L}$ as a function of $g$, with $\protect \alpha =1$ \ and $%
\protect \eta =0.8$. (a) symmetrical two-mode multi-PSS, (b) single mode multi-PSS.}
\end{figure}
\begin{figure}[tbh]
\label{Figure16} \centering%
\subfigure{
\begin{minipage}[b]{0.5\textwidth}
\includegraphics[width=0.83\textwidth]{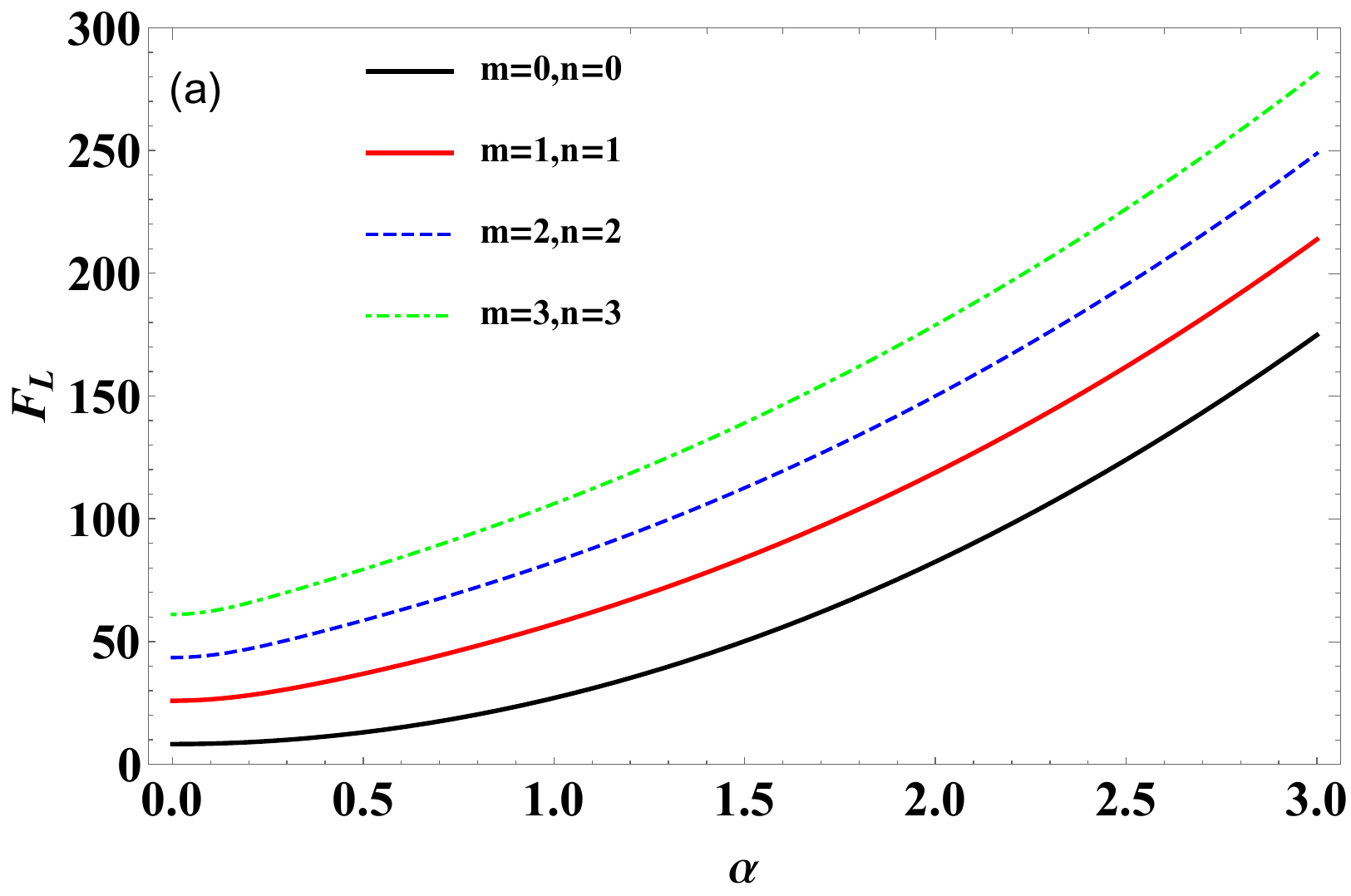}\\
\includegraphics[width=0.83\textwidth]{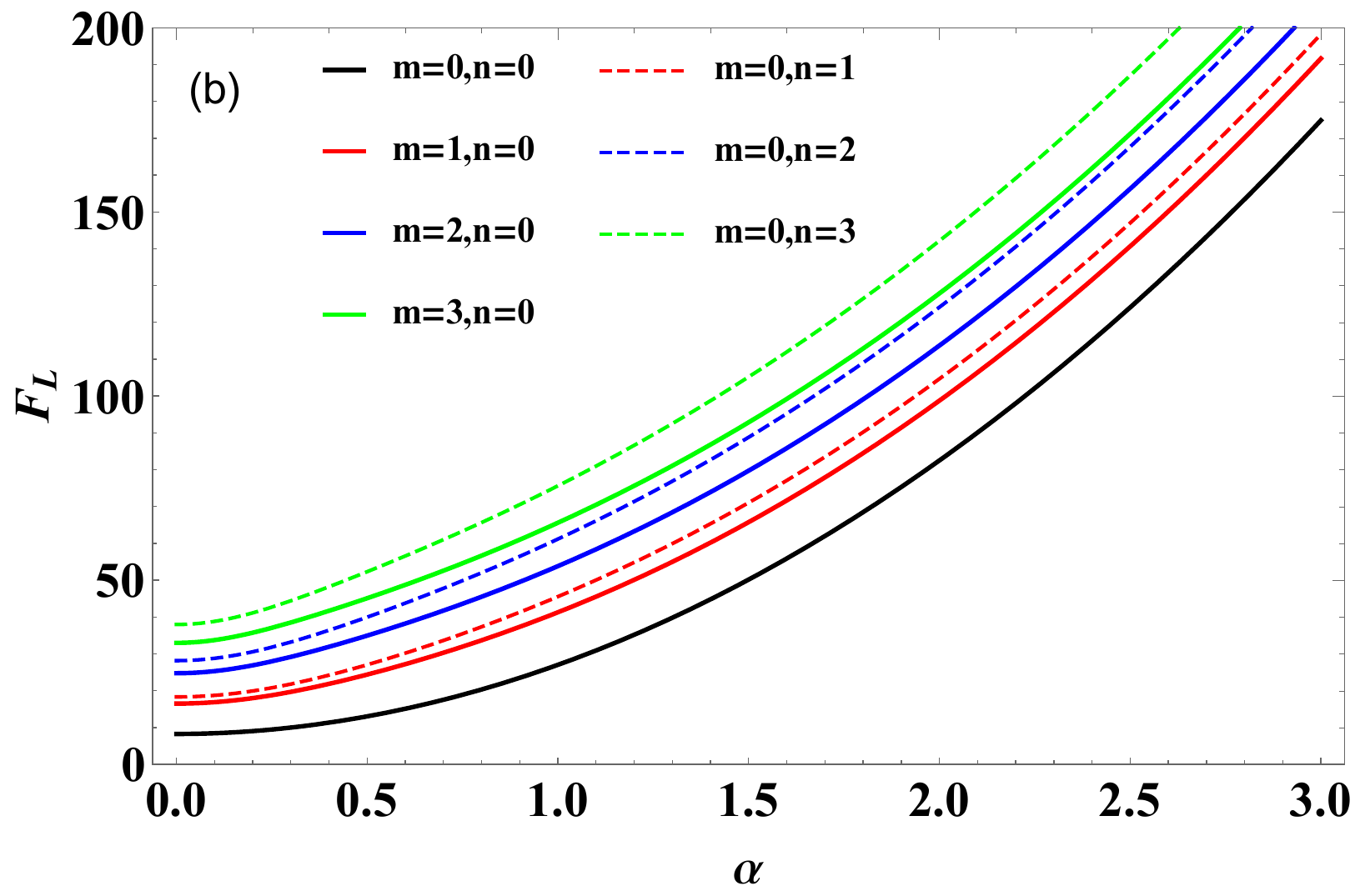}
\end{minipage}}
\caption{The $F_{L}$ as a function of $\protect \alpha $, with $g=1$ \ and $%
\protect \eta =0.8$. (a) symmetrical two-mode multi-PSS, (b) single mode multi-PSS.}
\end{figure}

Similar to the ideal case, the calculation of $\Delta \phi _{QCRB_{L}}$ is $%
\Delta \phi _{QCRB_{L}}=1/\sqrt{vF_{L}}$, with $v=1$. From Fig. 17, it is
evident that the $\Delta \phi _{QCRB_{L}}$ improves as the transmittance $%
\eta $ increases. For the multi-PSS, the $\Delta \phi _{QCRB_{L}}$ further
enhances with the number of subtracted photons. It can be observed that the
multi-PSS on mode $b$ exhibits a better $\Delta \phi _{QCRB_{L}}$ than that
on mode $a$ (Fig. 17(b)). Additionally, for both symmetrical and
asymmetrical multi-PSS, the improved $\Delta \phi _{QCRB_{L}}$ initially
increases and then decreases with the transmittance $\eta $.
\begin{figure}[tbh]
\label{Figure17} \centering%
\subfigure{
\begin{minipage}[b]{0.5\textwidth}
\includegraphics[width=0.83\textwidth]{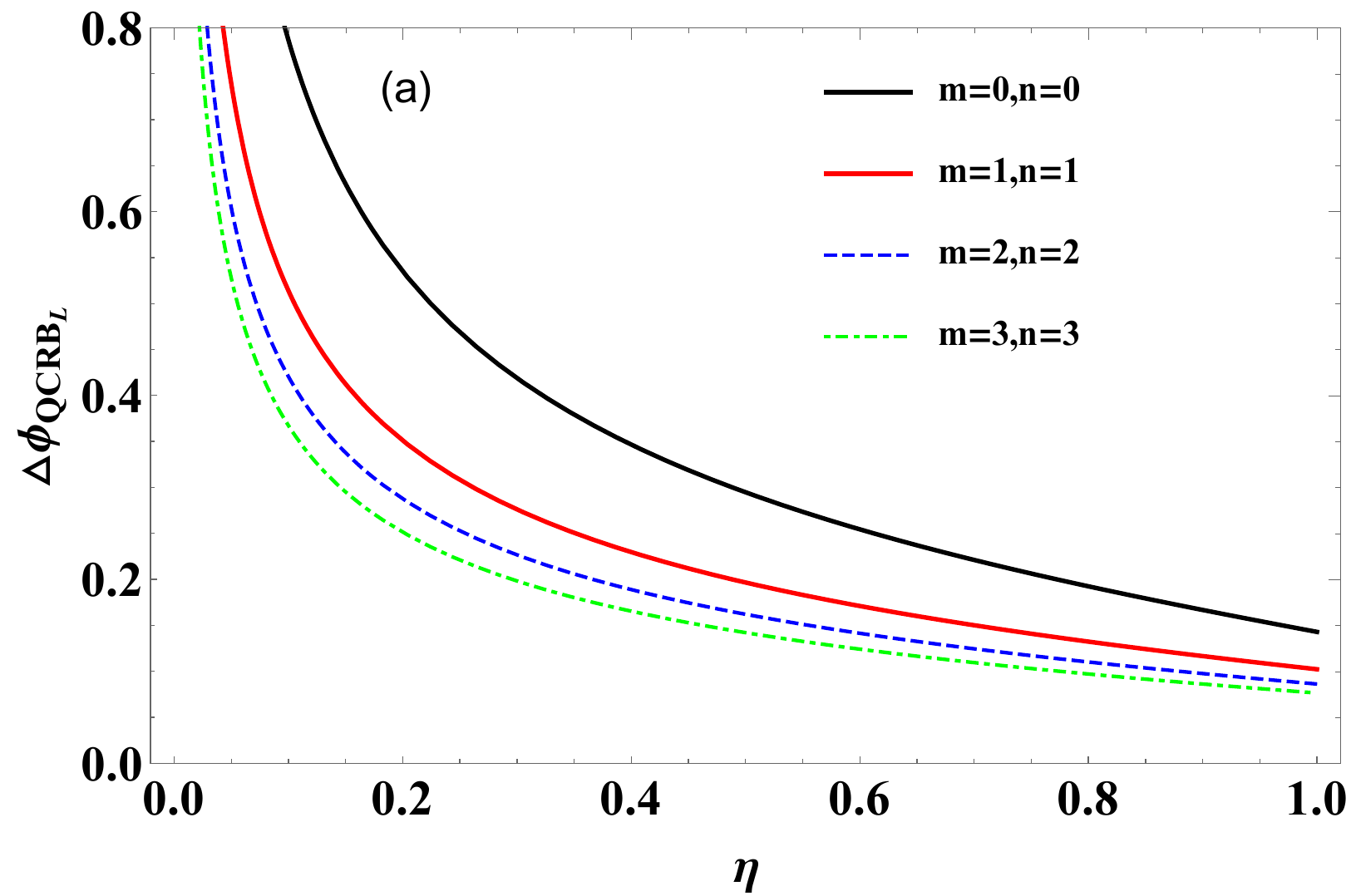}\\
\includegraphics[width=0.83\textwidth]{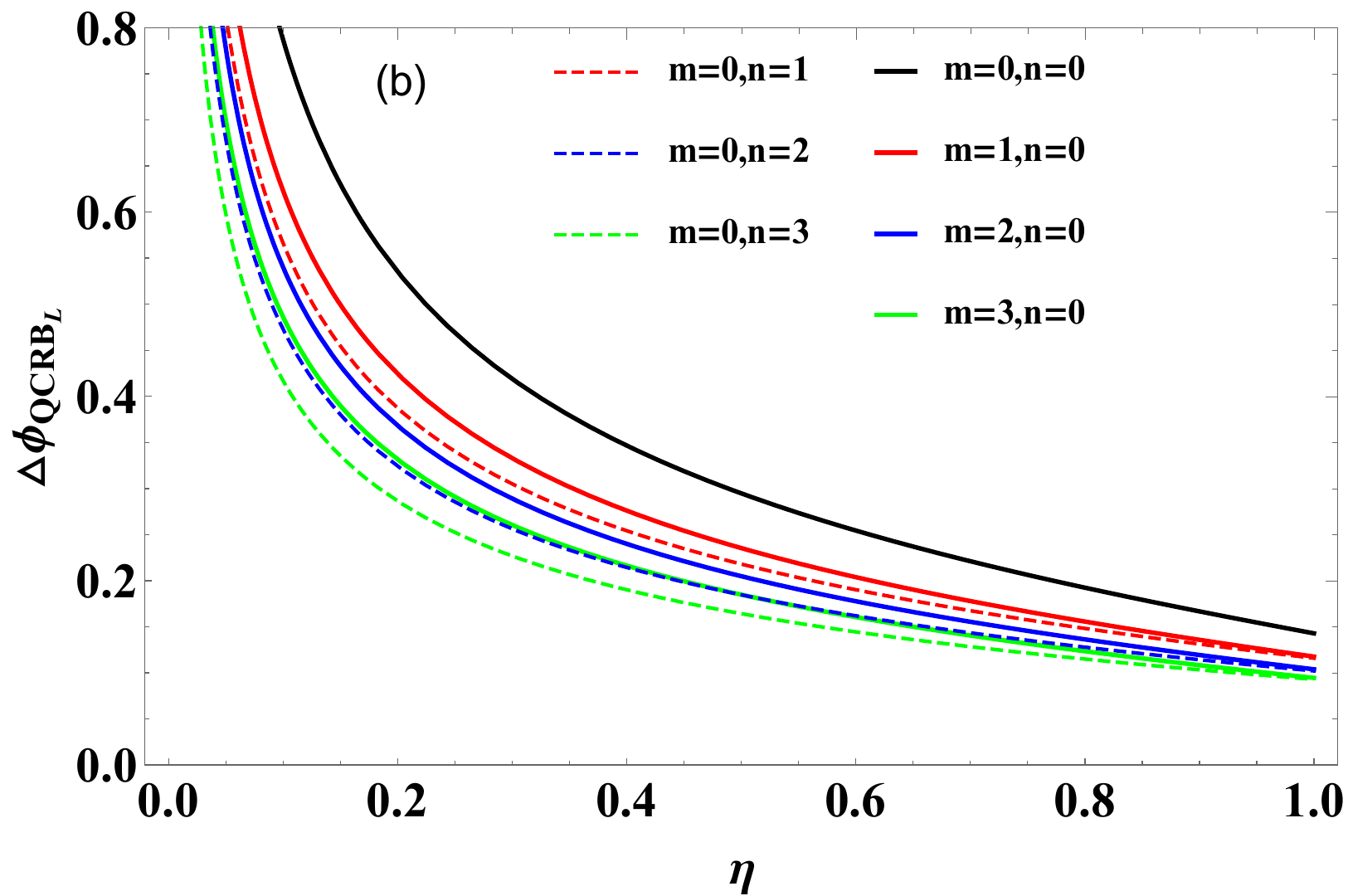}
\end{minipage}}
\caption{The $\Delta \protect \phi _{QCRB_{L}}$ as a function of
transmittance $\protect \eta $, with $g=1$ and $\protect \alpha =1$.
(a) symmetrical two-mode multi-PSS, (b) single mode multi-PSS.}
\end{figure}

\section{\protect \bigskip Conclusion}

In this paper, we have not only analyzed the effect of multi-PSS on the
phase sensitivity, the QFI and the QCRB in both ideal and real cases,
but also compared the results on different modes. Additionally, we have
investigated the effects of the gain coefficient $g$ of the parametric
amplifier, the coherent state amplitude $\alpha $ and the beam splitter
transmittance $T_{k}$, which simulates internal photon losses, on the system
performance. Through analytical comparison, we have confirmed that the
multi-PSS can improve the measurement accuracy of the SU(1,1) interferometer
and enhance the robustness to internal photon losses.

The results indicate that increasing the number of operated photons can
enhance the phase sensitivity $\Delta \phi $, and more operated photons
corresponding to better interferometer performance. The non-Gaussian
operation can increase the total mean photon number of the SU(1,1)
interferometer, consequently enhancing intramode correlations and quantum
entanglement between the two modes. Additionally, we analyze the effects of
performing separatively arbitrary photon subtraction on the two-mode inside
SU(1,1) interferometer, including the asymmetric properties of non-Gaussian
operations on the phase precision and the QFI. Regarding phase sensitivity,
multi-PSS on mode $a$ demonstrates superior performance overall under
certain parameters, attributed to homodyne detection on the mode $a$ of the
outport. Furthermore, in the presence of internal photon losses, the
multi-PSS exhibits better performance on mode $b$ when the losses are
substantial, while the opposite is true in the other case. In terms of the
QFI, the multi-PSS performs better on mode $b$ than on mode $a$.

In summary, the multi-PSS plays a role in overcoming the internal photon
losses in SU(1,1) interferometers and in improving the accuracy of quantum
measurements. This study underscores the potential of the multi-PSS as a
valuable tool for improving the performance of quantum metrology and
information processing systems.

\begin{acknowledgments}
This work is supported by the National Natural Science Foundation of China
(Grants No. 11964013 and No. 12104195) and the Training Program for Academic
and Technical Leaders of Major Disciplines in Jiangxi Province (No.
20204BCJL22053).
\end{acknowledgments}
\bigskip

\textbf{APPENDIX\ A : THE PHASE SENSITIVITY OF MULTI-PSS} \bigskip

In this Appendix, we give the calculation formulas of the phase sensitivity
for multi-PSS as follows
\begin{equation}
\Delta \phi =\frac{\sqrt{\left \langle \Psi _{out}^{1}\right \vert
X^{2}\left \vert \Psi _{out}^{1}\right \rangle -\left \langle \Psi
_{out}^{1}\right \vert X\left \vert \Psi _{out}^{1}\right \rangle ^{2}}}{%
|\partial \left \langle \Psi _{out}^{1}\right \vert X\left \vert \Psi
_{out}^{1}\right \rangle /\partial \phi |}.  \tag{A1}
\end{equation}
The output state $\left \vert \Psi _{out}^{1}\right \rangle $ is given by
equation (\ref{a2}) in our paper, so the expectations related to the phase
sensitivity for multi-PSS are specifically calculated as \cite{c0}
\begin{align}
& \left \langle \Psi _{out}^{1}\right \vert X\left \vert \Psi
_{out}^{1}\right \rangle   \notag \\
=& A^{2}[e^{-i\phi }\cosh gD_{m_{1}+1,n_{1},m_{2},n_{2}}e^{w_{1}}  \notag \\
& +\sinh gD_{m_{1},n_{1},m_{2},n_{2}+1}e^{w_{1}}  \notag \\
& +e^{i\phi }\cosh gD_{m_{1},n_{1},m_{2}+1,n_{2}}e^{w_{1}}  \notag \\
& +\sinh gD_{m_{1},n_{1}+1,m_{2},n_{2}}e^{w_{1}}],  \tag{A2}
\end{align}%
and
\begin{align*}
& \left \langle \Psi _{out}^{1}\right \vert X^{2}\left \vert \Psi
_{out}^{1}\right \rangle  \\
=& A^{2}[e^{-2i\phi }\cosh ^{2}gD_{m_{1}+2,n_{1},m_{2},n_{2}}e^{w_{1}} \\
& +e^{-i\phi }\sinh \left( 2g\right) D_{m_{1}+1,n_{1},m_{2},n_{2}+1}e^{w_{1}}
\\
& +\sinh ^{2}gD_{m_{1},n_{1},m_{2},n_{2}+2}e^{w_{1}} \\
& +e^{2i\phi }\cosh ^{2}gD_{m_{1},n_{1},m_{2}+2,n_{2}}e^{w_{1}} \\
& +e^{i\phi }\sinh \left( 2g\right) D_{m_{1},n_{1}+1,m_{2}+1,n_{2}}e^{w_{1}}
\\
& +\sinh ^{2}gD_{m_{1},n_{1}+2,m_{2},n_{2}}e^{w_{1}} \\
& +2\cosh ^{2}gD_{m_{1}+1,n_{1},m_{2}+1,n_{2}}e^{w_{1}} \\
& +e^{-i\phi }\sinh \left( 2g\right) D_{m_{1}+1,n_{1}+1,m_{2},n_{2}}e^{w_{1}}
\\
& +e^{i\phi }\sinh \left( 2g\right) D_{m_{1},n_{1},m_{2}+1,n_{2}+1}e^{w_{1}}
\\
& +2\sinh ^{2}gD_{m_{1},n_{1}+1,m_{2},n_{2}+1}e^{w_{1}}] \\
& +\cosh \left( 2g\right) .
\end{align*}

\bigskip
\textbf{APPENDIX\ B : THE QFI OF PHOTON LOSSES}\bigskip

The quantum state of the input state passing through the first OPA is
denoted as $\left \vert \psi \right \rangle =U_{S_{1}}\left \vert \alpha
\right \rangle _{a}\otimes \left \vert 0\right \rangle _{b},$ and its
corresponding density operator is represented as $\rho _{0}$, which
satisfies the following relation
\begin{equation}
Tr\rho _{0}=1.  \tag{B1}
\end{equation}%
The density matrix corresponding to the quantum state $U_{B_{a}}\left \vert
\psi \right \rangle $ of $\left \vert \psi \right \rangle $ after experiencing
photon losses is $\rho $
\begin{equation}
\rho =\sum_{l}\Pi _{l}\left( \eta \right) \rho _{0}\Pi _{l}^{\dagger }\left(
\eta \right) ,  \tag{B2}
\end{equation}%
where $\Pi _{l}\left( \eta \right) $ represents the Kraus operator, i.e.,
\begin{equation}
\Pi _{l}\left( \eta \right) =\sqrt{\frac{\left( 1-\eta \right) ^{l}}{l!}}%
\eta ^{\frac{n}{2}}a^{l}.  \tag{B3}
\end{equation}%
It satisfies
\begin{equation}
\text{Tr}\rho =1,\sum_{l}\Pi _{l}^{\dagger }\left( \eta \right) \Pi
_{l}\left( \eta \right) =1.  \tag{B4}
\end{equation}

The quantum state of $U_{B_{a}}\left \vert \psi \right \rangle $\ after photon
subtraction operation can be expressed as $A^{^{\prime
}}U_{P}U_{B_{a}}\left \vert \psi \right \rangle =A^{^{\prime
}}a^{m}b^{m}U_{B_{a}}\left \vert \psi \right \rangle $, where $A^{^{\prime }}$
is the normalization factor. It density operator $\rho ^{\prime }$ is
\begin{align}
\rho ^{\prime }& =A^{^{\prime }2}a^{m}b^{m}\rho b^{\dagger m}a^{\dagger m}
\notag \\
& =\sum_{l}\Pi _{l}\left( \eta \right) \rho _{0}^{\prime }\Pi _{l}^{\dagger
}\left( \eta \right) ,  \tag{B5}
\end{align}%
which satisfies Tr$\rho ^{\prime }=1$. Here $\rho _{0}^{\prime }=A^{^{\prime
}2}\eta ^{m}a^{m}b^{m}\rho _{0}b^{\dagger m}a^{\dagger m}=A^{^{\prime
}2}\eta ^{m}U_{P}\left \vert \psi \right \rangle \left \langle \psi \right \vert
U_{P}^{\dagger }=\left \vert \psi ^{\prime }\right \rangle \left \langle \psi
^{\prime }\right \vert $ is the density operator corresponding to the quantum
state $\left \vert \psi ^{\prime }\right \rangle $ after the input state
passes through the first OPA and then through photon subtraction, and $\rho
_{0}^{\prime }$ also satisfies $Tr\rho _{0}^{\prime }=1$. $A^{^{\prime
}}\eta ^{\frac{m}{2}}$ is the normalization factor of $\left \vert \psi
^{\prime }\right \rangle $
\begin{equation}
A^{^{\prime }2}\eta ^{m}=A_{0}^{2},  \tag{B6}
\end{equation}%
and $A_{0}$ reduces to $A$ for $T_{k}=1$. These derivations above utilize
the following equations%
\begin{equation}
\eta ^{\frac{n}{2}}a^{\dagger m}=\eta ^{\frac{m}{2}}a^{\dagger m}\eta ^{%
\frac{n}{2}},  \tag{B7}
\end{equation}%
\begin{equation}
a^{m}\eta ^{\frac{n}{2}}=\eta ^{\frac{m}{2}}\eta ^{\frac{n}{2}}a^{m}.
\tag{B8}
\end{equation}

The density operator $\rho ^{\prime }$ is equivalent to the density operator
of the quantum state obtained by first passing the input state through the
first OPA, then performing photon subtraction, and finally experiencing
photon losses (i.e. $\left \vert \psi ^{\prime }\right \rangle $, the quantum
state after photon losses). We consider the phase shifter operator $e^{i\phi
(a^{\dagger }a)}=e^{i\phi n}$ and introduce the parameter $\lambda $, which
accounts for the selection of whether photon losses occurs before or after
the phase shifter. Specifically, for $\lambda =0$, it corresponds to photon
losses occurring before the phase shifter, and for $\lambda =-1$, it
corresponds to photon losses occurring after the phase shifter.

Next, we define the Kraus operator
\begin{equation}
\Pi _{l}\left( \phi ,\eta ,\lambda \right) =\sqrt{\frac{\left( 1-\eta
\right) ^{l}}{l!}}e^{i\phi \left( n-\lambda l\right) }\eta ^{\frac{n}{2}%
}a^{l},  \tag{B9}
\end{equation}%
and it satisfies
\begin{equation}
\sum_{l}\Pi _{l}^{\dagger }\left( \phi ,\eta ,\lambda \right) \Pi _{l}\left(
\phi ,\eta ,\lambda \right) =1.  \tag{B10}
\end{equation}%
Hence, in the case of photon losses, the density operator corresponding to
the quantum state of $\left \vert \psi ^{\prime }\right \rangle $ after
passing through a phase shifter satisfies
\begin{equation}
\rho ^{\prime }\rightarrow \sum_{l}\Pi _{l}\left( \phi ,\eta ,\lambda
\right) \rho _{0}^{\prime }\Pi _{l}^{\dagger }\left( \phi ,\eta ,\lambda
\right).  \tag{B11}
\end{equation}

The consistency between the form of this density operator and the Kraus
operators $\Pi _{l}\left( \phi ,\eta ,\lambda \right) $, $\rho \left(
x\right) $ and $\Pi _{l}\left( x\right) $ in Ref. \cite{b12}, allows for the
utilization of the corresponding formulas.
\begin{equation}
F_{L}\leq C_{Q}=4\left[ \left \langle H_{1}\right \rangle -\left \langle
H_{2}\right \rangle ^{2}\right] ,  \tag{B12}
\end{equation}%
where
\begin{equation}
H_{1}=\sum_{l}\frac{d\Pi ^{\dagger }\left( \phi ,\eta ,\lambda \right) }{%
d\phi }\frac{d\Pi \left( \phi ,\eta ,\lambda \right) }{d\phi },  \tag{B13}
\end{equation}%
\begin{equation}
H_{2}=i\sum_{l}\frac{d\Pi ^{\dagger }\left( \phi ,\eta ,\lambda \right) }{%
d\phi }\Pi \left( \phi ,\eta ,\lambda \right) ,  \tag{B14}
\end{equation}%
and
\begin{align}
F_{L}& =\frac{4\eta \left \langle n_{0}\right \rangle \left \langle \Delta
n^{2}\right \rangle _{0}}{\left \langle \Delta n^{2}\right \rangle _{0}\left(
1-\eta \right) +\eta \left \langle n_{0}\right \rangle }  \notag \\
\ & =\frac{4F\eta \left \langle n_{0}\right \rangle }{\left( 1-\eta \right)
F+4\eta \left \langle n_{0}\right \rangle }.  \tag{B15}
\end{align}

For the multi-PSS with photon losses, the QFI can be obtained as
\begin{align}
F_{L}& =\frac{4F\eta \left \langle \psi ^{\prime }\right \vert a^{\dagger
}a\left \vert \psi ^{\prime }\right \rangle }{\left( 1-\eta \right) F+4\eta
\left \langle \psi ^{\prime }\right \vert a^{\dagger }a\left \vert \psi
^{\prime }\right \rangle }  \notag \\
\ & =\frac{4F\eta A_{0}^{2}D_{m_{1}+1,n_{1},m_{2}+1,n_{2}}e^{w_{1}}}{\left(
1-\eta \right) F+4\eta A_{0}^{2}D_{m_{1}+1,n_{1},m_{2}+1,n_{2}}e^{w_{1}}}.
\tag{B16}
\end{align}


\begin{thebibliography}{99}
\bibitem{01} A. N. Boto, P. Kok, D. S. Abrams, S. L. Braunstein, C. P.
Williams, and J. P. Dowling, Quantum interferometric optical lithography:
exploiting entanglement to beat the diffraction limit,\ Phys. Rev. Lett.
85(13), 2733 (2000).

\bibitem{02} J. J. Cooper, D. W. Hallwood, and J. A. Dunningham,
Entanglement-enhanced atomic gyroscope,\ Phys. Rev. A 81(4), 043624 (2010).

\bibitem{03} W. Wasilewski, K. Jensen, H. Krauter, J. J. Renema, M. V.
Balabas, and E. S. Polzik, Quantum noise limited and entanglement-assisted
magnetometry,\ Phys. Rev. Lett. 104(13), 133601 (2010).

\bibitem{04} F. Dolde et al., Electric-field sensing using single diamond
spins,\ Nat. Phys. 7(6), 459 (2011).

\bibitem{05} H. Muntinga et al., Interferometry with Bose-Einstein
condensates in microgravity,\ Phys. Rev. Lett. 110(9), 093602 (2013).

\bibitem{06} C. F. Ockeloen, R. Schmied, M. F. Riedel, and P. Treutlein,
Quantum metrology with a scanning probe atom interferometer,\ Phys. Rev.
Lett. 111(14), 143001 (2013).

\bibitem{07} K. Liu, C. Cai, J. Li, L. Ma, H. Sun, and J. Gao,
Squeezing-enhanced rotating-angle measurement beyond the quantum limit,\
Appl. Phys. Lett. 113(26), 261103 (2018).

\bibitem{08} Y. Zhai, Z. Yue, L. Li, and Y. Liu, Progress and applications
of quantum precision measurement based on SERF effect,\ Front. Phys. 10,
969129 (2022).

\bibitem{09} Y. Wu, J. Guo, X. Feng, L. Q. Chen, C. H. Yuan, and W. Zhang,
Atom-light Hybrid quantum gyroscope,\ Phys. Rev. Applied 14(6), 064023
(2020).

\bibitem{a1} B. Yurke, S. L. McCall, and J. R. Klauder, SU(2) and SU(1,1)
interferometers,\ Phys. Rev. A 33(6), 4033 (1986).

\bibitem{m1} J. Kong, F. Hudelist, Z. Y. Ou, and W. Zhang, Cancellation of
internal quantum noise of an amplifier by quantum correlation, Phys. Rev.
Lett. 111(3), 033608 (2013).

\bibitem{m3} C. M. Caves, Reframing SU(1,1) interferometry, Adv. Quantum
Technol. 3(11), 1900138 (2020).

\bibitem{2} J. Jing, C. Liu, Z. Zhou, Z. Y. Ou, and W. Zhang, Realization of
a nonlinear interferometer with parametric amplifiers,\ Appl. Phys. Lett.
99(1), 011110 (2011).

\bibitem{m4} J. Kong, Z. Y. Ou, and W. Zhang, Phase-measurement sensitivity
beyond the standard quantum limit in an interferometer consisting of a
parametric amplifier and a beam splitter, Phys. Rev. A 87(2), 023825 (2013).

\bibitem{m5} B. Chen, C. Qiu, S. Chen, J. Guo, L. Q. Chen, Z. Y. Ou, and W.
Zhang, Atom-light hybrid interferometer, Phys. Rev. Lett. 115(4), 043602
(2015).

\bibitem{m6} B. E. Anderson, P. Gupta, B. L. Schmittberger, T. Horrom, C.
Hermann-Avigliano, K. M. Jones, and P. D. Lett, Phase sensing beyond the
standard quantum limit with a variation on the SU(1, 1) interferometer,
Optica 4(7), 752--756 (2017).

\bibitem{m7} S. S. Szigeti, R. J. Lewis-Swan, and S. A. Haine, Pumped-up
SU(1, 1) interferometry, Phys. Rev. Lett. 118(15), 150401 (2017).

\bibitem{m9} G. Frascella, E. E. Mikhailov, N. Takanashi, R. V. Zakharov, O.
V. Tikhonova, and M. V. Chekhova, Wide-field SU(1,1) interferometer, Optica
6(9), 1233--1236 (2019).

\bibitem{m10} J. Liu, Y. Wang, M. Zhang, J. Wang, D. Wei, and H. Gao,
Ultra-sensitive phase measurement based on an SU(1, 1) interferometer
employing external resources and substract intensity detection, Opt. Express
28(26), 39443--39452 (2020).

\bibitem{m11} W. Du, J. F. Chen, Z. Y. Ou, and W. Zhang, Quantum dense
metrology by an SU(2)-in-SU(1, 1) nested interferometer, Appl. Phys. Lett.
117(2), 024003 (2020).

\bibitem{m12} D. Liao, J. Xin, and J. Jing, Nonlinear interferometer based
on two-port feedback nondegenerate optical parametricamplification, Opt.
Commun. 496, 127137 (2021).

\bibitem{3} A. M. Marino, N. V. Corzo Trejo, and P. D. Lett, Effect of
losses on the performance of an SU(1,1) interferometer,\ Phys. Rev. A 86(2),
023844 (2012).

\bibitem{4} D. Li, C. H. Yuan, Z. Y. Ou, and W. Zhang, The phase sensitivity
of an su(1, 1) interferometer with coherent and squeezed-vacuum light,\ New
J. Phys. 16(7), 073020 (2014).

\bibitem{5} M. Manceau, G. Leuchs, F. Khalili, and M. Chekhova, Detection
loss tolerant supersensitive phase measurement with an SU(1,1)
interferometer,\ Phys. Rev. Lett. 119(22), 223604 (2017).

\bibitem{m13} M. S. Kim, Recent developments in photon-level operations on
travelling light fields, J. Phys. B: At., Mol. Opt. Phys. 41(13), 133001
(2008).

\bibitem{m14} V. Parigi, A. Zavatta, M. Kim, and M. Bellini, Probing quantum
commutation rules by addition and subtraction of single photons to/from a
light field,\ Science 317(5846), 1890--1893 (2007).

\bibitem{m15} M. Bellini and A. Zavatta, Manipulating light states by
single-photon addition and subtraction,\ Prog. Opt. 55, 41--83 (2010).

\bibitem{10} P. S. Yan, L. Zhou, W. Zhong, and Y. B. Sheng, Advances in
quantum entanglement purification,\ Sci. China-Phys. Mech. Astron. 66(5),
250301 (2023).

\bibitem{a21} C. H. Bennett, H. J. Bernstein, S. Popescu, and B. Schumacher,
Concentrating partial entanglement by local operations,\ Phys. Rev. A 53(4),
2046 (1996).

\bibitem{a22} R. Horodecki, P. Horodecki, M. Horodecki, and K. Horodecki,
Quantum entanglement,\ Rev. Mod. Phys. 81(2), 865 (2009).

\bibitem{a23} S. L. Braunstein and P. van Loock, Quantum information with
continuous variables,\ Rev. Mod. Phys. 77(2), 513 (2005).

\bibitem{m16} P. G. Kwiat, S. Barraza-Lopez, A. Stefanov, and N. Gisin,
Experimental entanglement distillation and 'hidden' non-locality, Nature
409(6823), 1014--1017 (2001).

\bibitem{m17} L. L. Guo, Y. F. Yu, and Z. M. Zhang, Improving the phase
sensitivity of an SU(1,1) interferometer with photon-added squeezed vacuum
light,\ Opt. Express 26(22), 29099 (2018).

\bibitem{m18} K. Zhang, Y. H. Lv, Y. Guo, J. T. Jing, and W. M. Liu,
Enhancing the precision of a phase measurement through phase-sensitive
non-Gaussianity,\ Phys. Rev. A 105(4), 042607 (2022).

\bibitem{m19} Y. K. Xu, S. K. Chang, C. J. Liu, L. Y. Hu, and S. Q. Liu,
Phase estimation of an SU(1,1) interferometer with a coherent superposition
squeezed vacuum in a realistic case,\ Opt. Express 30(21), 38178 (2022).

\bibitem{m20} A. Zavatta, V. Parigi, M. S. Kim, H. Jeong, and M. Bellini,
Experimental demonstration of the Bosonic commutation relation via
superpositions of quantum operations on thermal light fields,\ Phys. Rev.
Lett. 103(14), 140406 (2009).

\bibitem{b1} M. V. Chekhova and Z. Y. Ou, Nonlinear interferometers in
quantum optics,\ Adv. Opt. Photonics 8(1), 252890 (2016).

\bibitem{b2} H. Ma and Y. Liu, Super-resolution localization microscopy:
Toward high throughput, high quality, and low cost,\ APL Photonics 5(6),
080902 (2020).

\bibitem{7} Z. Y. Ou, Enhancement of the phase-measurement sensitivity
beyond the standard quantum limit by a nonlinear interferometer,\ Phys. Rev.
A 85(2), 023815 (2012).

\bibitem{9} J. Xin, Phase sensitivity enhancement for the SU(1,1)
interferometer using photon level operations,\ Opt. Express 29(26),
43970-43984 (2021).

\bibitem{b3} C. W. Helstrom, Minimum mean-squared error of estimates in
quantum statistics, Phys. Lett. A 25(2), 101 (1967).

\bibitem{b4} C. W. Helstrom, Quantum detection and estimation theory, J.
Stat. Phys. 1(2), 231 (1969).

\bibitem{c0} Y. Xu, T. Zhao, Q. Kang, C. Liu, L. Hu, and S. Liu, Phase
sensitivity of an SU(1,1) interferometer in photon-loss via photon
operations,\ Opt. Express 31(5), 8414 (2023).

\bibitem{c1} M. Xiao, L. A. Wu, and H. J. Kimble, Precision measurement
beyond the shot-noise limit,\ Phys. Rev. Lett. 59(3), 278 (1987).

\bibitem{c2} R. Demkowicz-Dobrza\'{n}ski, M. Jarzyna, and J. Ko\l ody\'{n}%
ski, Quantum limits in optical interferometry,\ Prog. Optics 60, 345-435
(2015).

\bibitem{c3} M. Bradshaw, P. K. Lam, and S. M. Assad, Ultimate precision of
joint quadrature parameter estimation with a Gaussian probe,\ Phys. Rev. A
97(1), 012106 (2018).

\bibitem{b5} D. Li, C. H. Yuan, Z. Y. Ou, and W. Zhang, The phase
sensitivity of an SU(1,1) interferometer with coherent and squeezed-vacuum
light,\ New J. Phys 16(7), 073020 (2014).

\bibitem{b6} X. Y. Hu, C. P. Wei, Y. F. Yu, and Z. M. Zhang, Enhanced phase
sensitivity of an SU(1,1) interferometer with displaced squeezed vacuum
light,\ Front. Phys. 11(3), 114203 (2016).

\bibitem{b7} P. M. Anisimov, G. M. Raterman, A. Chiruvelli, W. N. Plick, S.
D. Huver, H. Lee, and J. P. Dowling, Quantum metrology with two-mode
squeezed vacuum: parity detection beats the Heisenberg limit,\ Phys. Rev.
Lett. 104(10), 103602 (2010).

\bibitem{b8} D. Li, B. T. Gard, Y. Gao, C. H. Yuan, W. Zhang, H. Lee, and J.
P. Dowling, Phase sensitivity at the Heisenberg limit in an SU(1,1)
interferometer via parity detection,\ Phys. Rev. A 94(6), 063840 (2016).

\bibitem{b10} S. Ataman, A. Preda, and R. Ionicioiu, Phase sensitivity of a
Mach-Zehnder interferometer with single-intensity and difference-intensity
detection,\ Phys. Rev. A 98(4), 043856 (2018).

\bibitem{b11} D. Li, C. H. Yuan, Y. Yao, W. Jiang, M. Li, and W. Zhang,
Effects of loss on the phase sensitivity with parity detection in an SU(1,1)
interferometer,\ J. Opt. Soc. Am. B 35(5), 309106 (2018).

\bibitem{d1} C. M. Caves, Quantum-mechanical noise in an interferometer,\
Phys. Rev. D 23(8), 1693 (1981).

\bibitem{d2} O. Assaf and Y. Ben-Aryeh, Quantum mechanical noise in
coherent-state and squeezed-state Michelson interferometers,\ J. Opt. B:
Quantum Semiclass. Opt. 4(1), 49 (2002).

\bibitem{d3} J. Beltran and A. Luis, Breaking the Heisenberg limit with
inefficient detectors,\ Phys. Rev. A 72(4),045801 (2005).

\bibitem{b12} B. M. Escher, R. L. de Matos Filho, and L. Davidovich, General
framework for estimating the ultimate precision limit in noisy
quantum-enhanced metrology,\ Nat. Phys. 7(5), 406 (2011).

\bibitem{b13} Helstrom C W, Quantum detection and estimation theory
(Academic, 1976), 123.

\bibitem{b14} S. K. Chang, W. Ye, H. Zhang, L. Y. Hu, J. H. Huang, and S. Q.
Liu, Improvement of phase sensitivity in an SU(1, 1) interferometer via a
phase shift induced by a Kerr medium,\ Phys. Rev. A 105(3), 033704 (2022).

\bibitem{b15} S. K. Chang, C. P. Wei, H. Zhang, Y. Xia, W. Ye, and L. Y. Hu,
Enhanced phase sensitivity with a nonconventional interferometer and
nonlinear phase shifter,\ Phys. Lett. A 384(29), 126755 (2020).
\end{thebibliography}
\end{document}